\documentclass[twocolumn,pra,twocolumn,superscriptaddress]{revtex4}
\usepackage{color}
\usepackage{amsmath}
\usepackage{amssymb}
\usepackage{graphicx}
\usepackage{tikz}
\usepackage{lipsum}
\usepackage{yfonts}

\usepackage{float}
\usepackage{bbm}

\usepackage{epsfig}
\usepackage{array}

\usepackage[unicode=true,
bookmarks=true,bookmarksnumbered=false,bookmarksopen=false,
breaklinks=false,pdfborder={0 0 1},backref=false,colorlinks=true]
{hyperref}
\hypersetup{
	linkcolor=magenta, urlcolor=blue, citecolor=blue, pdfstartview={FitH}, hyperfootnotes=false, unicode=true}

\makeatletter
\@ifundefined{textcolor}{}
{%
	\definecolor{BLACK}{gray}{0}
	\definecolor{WHITE}{gray}{1}
	\definecolor{RED}{rgb}{1,0,0}
	\definecolor{GREEN}{rgb}{0,1,0}
	\definecolor{BLUE}{rgb}{0,0,1}
	\definecolor{CYAN}{cmyk}{1,0,0,0}
	\definecolor{MAGENTA}{cmyk}{0,1,0,0}
	\definecolor{YELLOW}{cmyk}{0,0,1,0}
}

\usepackage{amsfonts}
\usepackage{tabularx}
\usepackage{dcolumn}
\usepackage{bm}
\usepackage{graphicx}
\usepackage{epstopdf}
\usepackage{times}
\hypersetup{urlcolor=blue}

\usepackage{cancel}
\usepackage{ulem}

\makeatother

\newcommand\blue[1]{{\color{black}#1}}
\newcolumntype{C}[1]{>{\centering\arraybackslash$}p{#1}<{$}}

\begin{document}




\title{\blue{Coupling two charge qubits via a superconducting resonator operating in the resonant and dispersive regimes}}

\author{Chengxian Zhang}
\affiliation{School of Physical Science and Technology, Guangxi University, Nanning 530004, China}

\author{Guo Xuan Chan}
\affiliation{Department of Physics, City University of Hong Kong, Tat Chee Avenue, Kowloon, Hong Kong SAR, China}
\affiliation{City University of Hong Kong Shenzhen Research Institute, Shenzhen  518057, China}

\author{Xin Wang}\email{x.wang@cityu.edu.hk}
\affiliation{Department of Physics, City University of Hong Kong, Tat Chee Avenue, Kowloon, Hong Kong SAR, China}
\affiliation{City University of Hong Kong Shenzhen Research Institute, Shenzhen  518057, China}
\author{Zheng-Yuan Xue}  \email{zyxue83@163.com}
\affiliation{Guangdong Provincial Key Laboratory of Quantum Engineering and Quantum Materials,
	and School of Physics\\ and Telecommunication Engineering, South China Normal University, Guangzhou  510006, China}
\affiliation{Guangdong-Hong Kong Joint Laboratory of Quantum Matter, and Frontier Research Institute for Physics,\\ South China Normal University, Guangzhou  510006, China}

\date{\today}

\begin{abstract}

A key challenge for semiconductor quantum-dot charge qubits is the realization of long-range qubit coupling and performing high-fidelity gates based on it. Here, we describe a new type of charge qubit formed by an electron confined in a triple-quantum-dot system, enabling single and two-qubit gates working in the dipolar and quadrupolar detuning sweet spots. We further present the form for the long-range dipolar coupling between the charge qubit and the superconducting resonator. Based on the hybrid system composed of the qubits and the resonator, we present two types of entangling gates: the dynamical iSWAP gate and holonomic entangling gate, which are operating in the dispersive and resonant regimes, respectively. We find that the fidelity for the iSWAP gate can reach fidelity higher than 99\% for the noise level typical in experiments. Meanwhile, the fidelity for the holonomic gate can surpass 98\% if the anharmonicity in the resonator is large enough. Our proposal offers an alternative useful way to build up high-fidelity quantum computation for charge qubits in semiconductor quantum dot.

\end{abstract}

\maketitle

\section{Introduction}
\blue{The charge qubit in semiconductor quantum dots \cite{Shinkai.09,Petersson.10,Cao.13,Li.15,Kim.15, Ward.16,Yang.19b} is a promising candidate to realize universal quantum computing due to its all-electrical control and fast gate operation. Normally, the gate duration for the charge qubits can be as fast as several nanoseconds thanks to the large tunneling and detuning between the neighboring dots. Although recent experiments \cite{Noiri.22,Xue.22,Madzik.22} for spin qubits in silicon has reported fidelity exceeding 99\% for both single and two-qubit gates, the gating time there can be as long as $\mu s$ due to the small value of the microwave-driven Rabi frequency. Therefore, the charge qubits have potential advantages warranting further studies. On the other hand,} it suffers heavily from the charge noise \cite{Dial.13}, resulting in rather short coherent time and thus low gate fidelity \cite{Kim.15}. Despite the progress over the past years, the two-qubit quantum gate-fidelity in the experiment remains below 90\% \cite{Kim.15}, which motivates us to search further useful methods to design new types of charge qubits, aiming at mitigating the gate-fidelity.

As the isolated qubits are scaling up, how to implement distant and high-fidelity entangling gate between the neighbouring qubits remains another challenge. Typically, qubit-qubit interaction can be implemented for two charge qubits via direct capacitive coupling between two double quantum dots (DQDs), where the interaction range is only about 100 nm \cite{Van.18}. With this capacitive coupling, the entangling gate can only achieve gate-fidelity lower than 70\% \cite{Li.15,macquarrie.20}. On the other hand, the electrons confined in the quantum dots can form relatively large dipole moment when the dots are detuned, due to the delocalized wave-function. Therefore, the charge qubits has great potential to be coupled to the superconducting transmission-line resonator using the dipole moment. Recent experiment has demonstrated that both resonant (real) and nonresonant (virtual) resonator-mediated coherent interactions between two separated DQDs are possible \cite{Van.18}. The range of interaction between two charge qubits there can be substantially increased up to several tens of micrometers \cite{Van.18}.

Recently, it is found that one electron confined in a linear triple-quantum dot (TQD) can also be used to encode the so-called charge quadrupole (CQ) qubit \cite{Friesen.17}, which can benefit from the decoherence-free subspace and the dipolar sweet spot, where the dipolar detuning fluctuation is minimized. By using this quadrupole moment of the electron rather than the dipole moment, the long-range coupling between two CQ qubits and the resonator is experimentally realized \cite{koski.20}. Although the CQ qubit can work in the decoherence-free subspace, it still confronts severe leakage induced by the charge noise. To mitigate this leakage, composite pulses are required \cite{Ghosh.17}, which however prolongs the gate time.

Inspired by the CQ qubit, we find that (as shown below) one electron confined in a TQD can alternatively form another type of charge qubit. The logical basis states are defined as the two-lowest eigenstates, which are different from the one for the CQ qubit. Except for the diplolar sweet spot, the charge qubit considered here can also benefit from the quadrupolar sweet spot, where the leading order of the quadrupolar detuning fluctuation is eliminated.

Here, we investigate how the two spatially separated charge qubits defined in the TQD can be entangled with each other via dipolar coupling to a \blue{superconducting} resonator \cite{blais.07,Srinivasa.16, scarlino.19,landig.19}. The resonator field is coupled to the variation of the dipolar detuning of the qubit such that the oscillation in the dipolar detuning can be controlled by the resonator voltage. We have derived the specific form of the coupling between the qubit and the resonator. We further estimate the coupling strength considering the present experimental parameters for the TQD and the resonator. We will present two approaches to construct the entangling gates. When each qubit is in resonance with the resonator, one is able to achieve a holonomic entangling gate. While this hybrid system is working in the dispersive regime, an iSWAP gate is obtained. We numerically simulate  the fidelity for these two entangling gates considering with the present experimental decoherence parameters. \blue{We surprisingly find that the fidelity for the iSWAP gate can surpass 99\%, considering the noise level in experiments. While the fidelity for the holonomic gate sensitively depends on the anharmonicity in the resonator. }

\begin{figure}
\includegraphics[width=0.95\columnwidth]{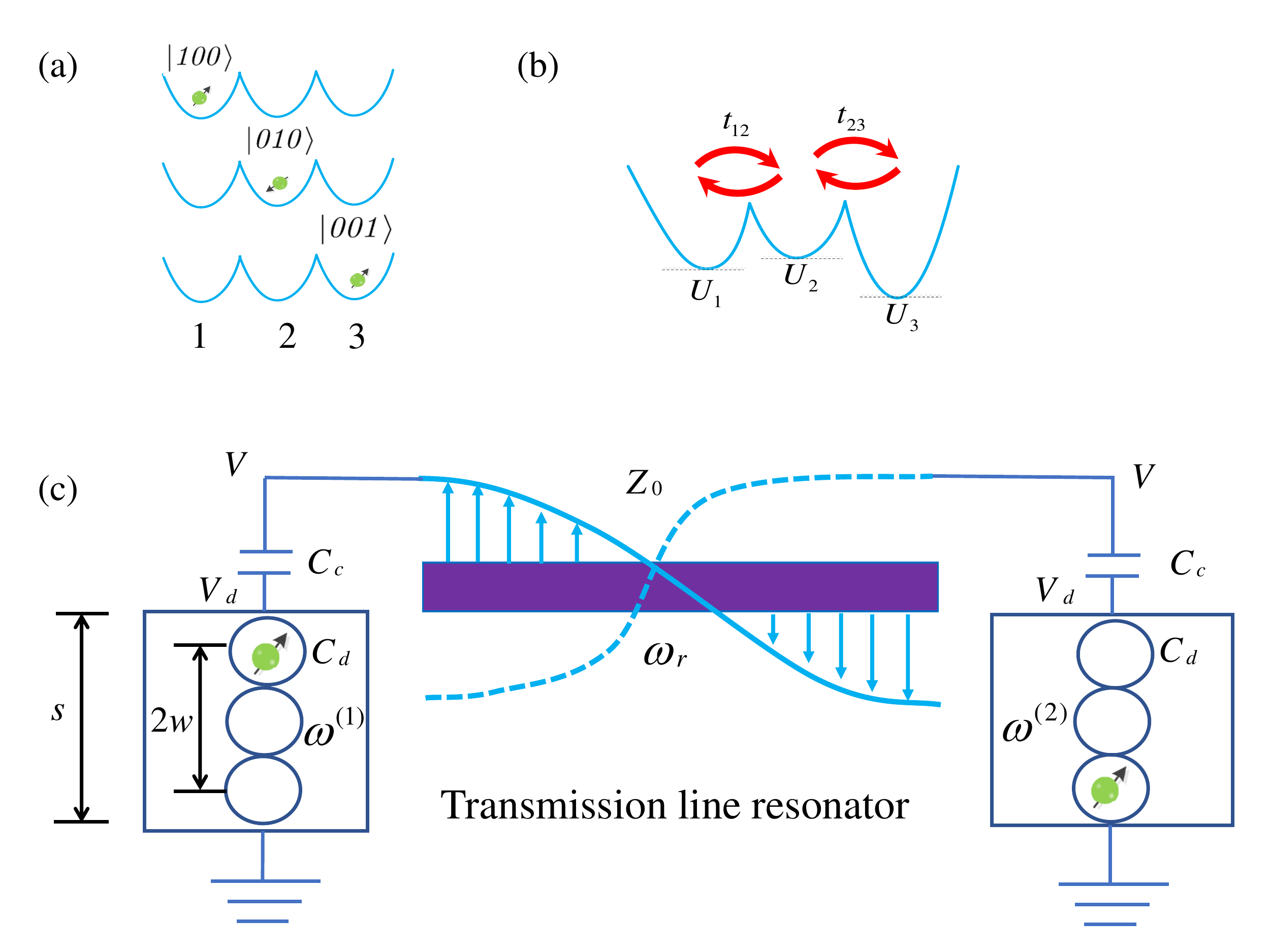}
\caption{Schematic illustration of the HC qubit and the coupling between the TQDs and the resonator. (a) The quantum dots are labeled by 1, 2 and 3 from the left to the right, which corresponds to the position basis states $|\mathit{100}\rangle$, $|\mathit{010}\rangle$ and $|\mathit{001}\rangle$. (b) The site potential for each dot and the tunneling between neighboring dots are depicted. (c) The coupling between the resonator and the TQDs, including their geometry and the parameters used to determine the coupling strength $g$.
}
\label{fig:quantumdot}
\end{figure}

\section{Double sweet spots in the TQD}\label{sec:model}

\begin{figure}
	\includegraphics[width=0.9\columnwidth]{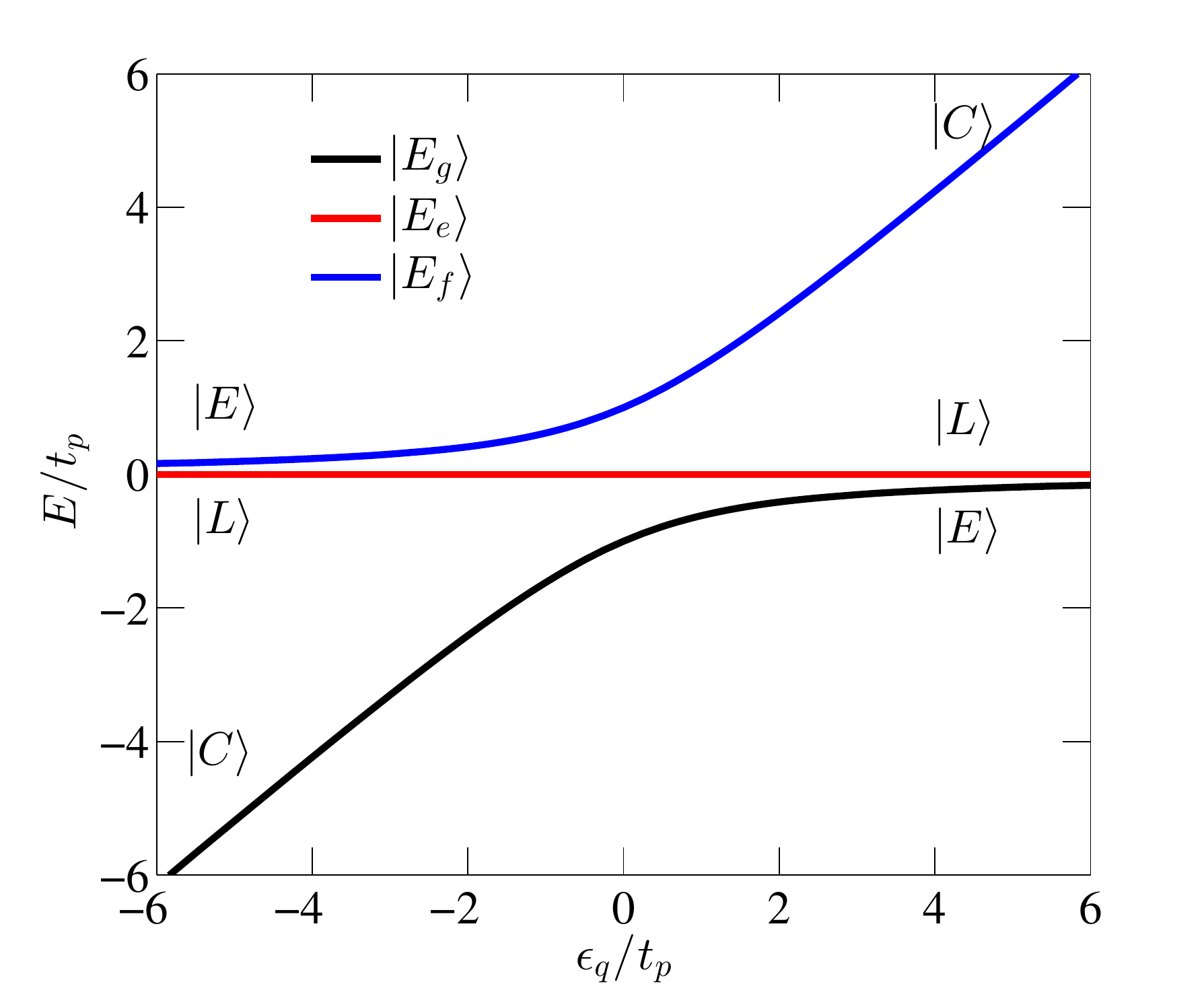}
	\caption{Energy level of the triple-quantum-dot system. The parameters are set to be $\bar{\epsilon}_{d}=t_{m}=0$. The ground, first excited and second excited states are denoted as $|g\rangle$, $|e\rangle$ and $|f\rangle$, which corresponds to the energy level $|E_{g}\rangle$, $|E_{e}\rangle$ and $|E_{f}\rangle$.}
	\label{fig:energy}
\end{figure}

As shown in Fig.~\ref{fig:quantumdot}(a), a single electron confined in a linear TQD can occupy the left, middle, or right dots, which corresponds to position states labeled by $\left|\mathit{100}\right\rangle$, $\left|\mathit{010}\right\rangle$, and $\left|\mathit{001}\right\rangle$, respectively.
The Hamiltonian in the position bases is \cite{Friesen.17}
\begin{equation}
\mathcal{H}^{(0)}=\begin{aligned}
\left( {\begin{array}{*{20}{c}}
	\epsilon_{d}&t_{12}&0\\
	t_{12}&\epsilon_{q}&t_{23}\\
	0&t_{23}&-\epsilon_{d}\\
	\end{array}} \right)
\end{aligned}
\label{eq:Hpo}
\end{equation}
Here, $t_{12}$ and $t_{23}$ are tunnel couplings between adjacent dots as indicated in Fig.~\ref{fig:quantumdot}(b). $\epsilon_{d}=(U_{1}-U_{3})/2$ and $\epsilon_{q}=U_{2}-(U_{1}+U_{3})/2$ are defined as the dipolar and quadrupolar detuning, respectively. Here, $U_{i}$ ($i$=1,2,3) denotes the site energy for the $i$th quantum dot. We note that, all parameters concerned here are real numbers, and we take $\hbar=1$ for simplicity. Each element in $\mathcal{H}^{(0)}$ can be controlled independently via the gate voltages \cite{Friesen.17}. In the ``even-odd'' bases,  i.e.~$\{|E\rangle=(|\mathit{100}\rangle+|\mathit{001}\rangle) / \sqrt{2},|C\rangle=|\mathit{010}\rangle,|L\rangle=(|\mathit{100}\rangle-|\mathit{001}\rangle) / \sqrt{2}\}$, $\mathcal{H}^{(0)}$ can be transformed  to
\begin{equation}
H=\begin{aligned}
\left( {\begin{array}{*{20}{c}}
	0&t_{p}&\epsilon_{d}\\
	t_{p}&\epsilon_{q}&t_{m}\\
	\epsilon_{d}&t_{m}&0\\
	\end{array}} \right)
\end{aligned}
\label{eq:Hpeel}
\end{equation}
where $t_{p}=(t_{12}+t_{23})/\sqrt{2}$, $t_{m}=(t_{12}-t_{23})/\sqrt{2}$.

\begin{figure*}
	\includegraphics[width=2\columnwidth]{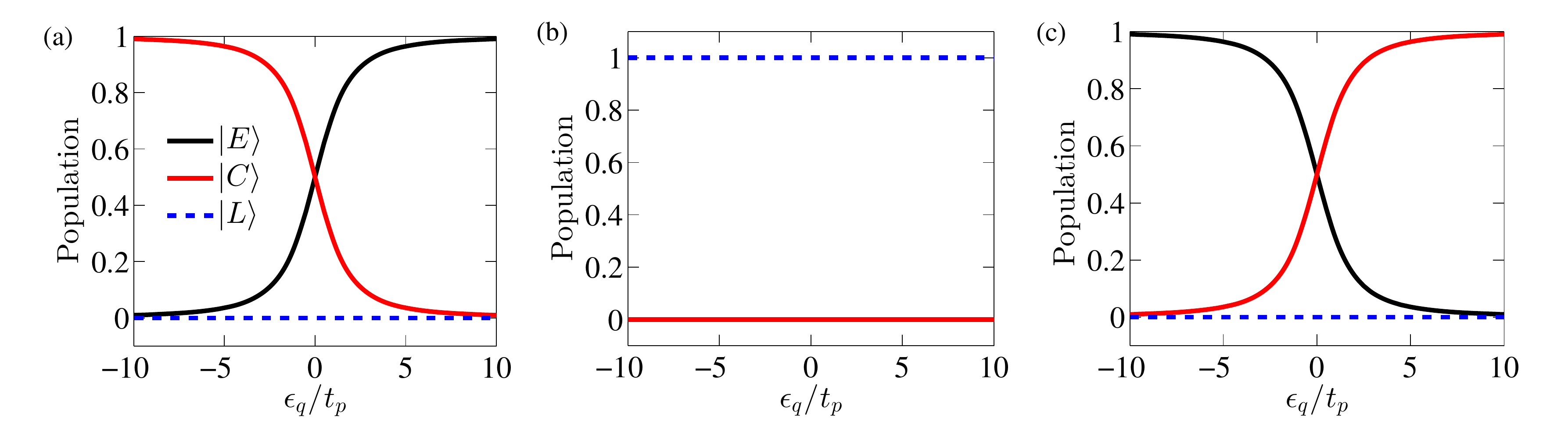}
	\caption{Population of the triple-quantum-dot system for the ground (a), first excited (b) and second excited states (c). The qubit operation is preferred to be in the region $\epsilon_{q}\gg t_{p}$, where the population of the ground state is dominated by $|E\rangle$, while the first excited state is mainly the state $|L\rangle$. The effective coupling between the qubit and the resonator is maximized in this region (as shown in the section.~\ref{sec:coupling}). The parameters are set to be $\bar{\epsilon}_{d}=t_{m}=0$.}
	\label{fig:popu}
\end{figure*}

Charge noise can cause fluctuations for both dipolar and quadrupolar detunings. We model the noise as $\epsilon_{d}\rightarrow\bar{\epsilon}_{d}+\delta\epsilon_{d}$ and $\epsilon_{q}\rightarrow\bar{\epsilon}_{q}+\delta\epsilon_{q}$, where $\bar{\epsilon}_{d}$ and $\bar{\epsilon}_{q}$ are the mean values, and $\delta\epsilon_{d}$ and $\delta\epsilon_{q}$ are the dipolar and quadrupolar fluctuations \cite{Friesen.17}. In the following, we consider the symmetric operating for the dipolar detuning i.e., $\bar{\epsilon}_{d}=0$.  We assume that the noise is quasi-static, so that $\delta\epsilon_{d}$ and $\delta\epsilon_{q}$ are treated as constants. This approximation is justified by the fact that the charge noises typically vary on a time scale of about 100 $\rm{\mu s}$ \cite{Wang.14,Ghosh.17}, much longer than the gating time (on the scale of ns) for a charge qubit. We then separate the Hamiltonian $H$ into two parts $H_0=H(\delta\epsilon_{d}=\delta\epsilon_{d}=0)$ and $H'=H-H_0$. $H'$ refers to the fluctuation components, while $H_0$ can be diagonalized analytically as seen in Appendix~\ref{appx:eigen}. The eigenstates are defined as $|g\rangle$, $|e\rangle$ and $|f\rangle$, which denote the ground, the first and second excited states respectively. Their corresponding eigenvalues are $E_{g}$, $E_{e}$ and $E_{f}$. Assuming $|\delta\epsilon_{d}|, |\delta\epsilon_{d}|\ll t_{p}, t_{m}$, we can expand the qubit excitation energy $\omega_{ge}=E_{ge}=E_{e}-E_{g}$ as

\begin{equation}
\begin{aligned}
E_{ge}&=\frac{1}{2}\left(\sqrt{4\left(t_{p}^{2}+t_{m}^{2}\right)+\bar{\epsilon}_{q}^{2}}-\bar{\epsilon}_{q}\right) \\
&- \frac{t_{p} t_{m} \left(3+\frac{\bar{\epsilon}_{q}}{\sqrt{4\left(t_{p}^{2}+t_{m}^{2}\right)+\bar{\epsilon}_{q}^{2}}}\right)}{t_{p}^{2}+t_{m}^{2}} \delta\epsilon_{d}\\ 
&+\frac{1}{2}\left(\frac{\bar{\epsilon}_{q}}{\sqrt{4\left(t_{p}^{2}+t_{m}^{2}\right)+\bar{\epsilon}_{q}^{2}}}-1\right) \delta\epsilon_{q} \\
&+ O\left((\delta \epsilon_{q}+\delta\epsilon_{d})^{2}\right).
\end{aligned}
\label{eq:expand}
\end{equation}
It is clear that the first term on the right hand side of Eq.~(\ref{eq:expand}) represents the qubit excitation energy without noise, the second and third terms relate to the dipolar and quadrupolar fluctuations. Setting $t_{m}=0$, i.e., $t_{12}=t_{23}$, one is able to find a dipolar detuning sweet spot ($\partial{E_{ge}}/\partial{\epsilon_{d}}|_{t_{m}=0}=0$).  In the assumption of $t_{m}=0$, one can further obtain another quadrupole detuning sweet spot ($\partial{E_{ge}}/\partial{\epsilon_{q}}=0$) when it satisfies $\bar{\epsilon}_{q}\gg t_{p}$.  Although the dipolar detuning sweet spot has been widely studied \cite{Friesen.17,Kratochwil.21}, this quadrupole detuning sweet spot of the charge qubit in TQD is not yet reported. To take full advantage of these double sweet spots, in this work we are considering the operating region $\bar{\epsilon}_{d}=t_{m}=0$ and $\bar{\epsilon}_{q}\gg t_{p}$. \blue{Note that, although the two tunnel couplings in the experiments can be adjusted conveniently by a barrier gate \cite{Russ.18}, they cannot be totally identical due to either imperfect control of the gate or charge noise fluctuation on the tunneling. Nevertheless, by modeling  $t_{p}\rightarrow\bar{t}_{p}+\delta t_{p}$ and $t_{m}\rightarrow\bar{t}_{m}+\delta t_{m}$, one finds that the fluctuation on the dipolar detuning is approximated to be $\frac{4 \delta t_{m}}{t_{p}+\delta t_{p}} \delta \epsilon_{d}$. Here, $\bar{t}_{p}$ and $\bar{t}_{m}$ are the mean values, while $\delta t_{p}$ and $\delta t_{m}$ are the related fluctuations. Since $\delta t_{m}, \delta t_{p} \ll t_{p}, \delta \varepsilon_{d}$ \cite{Friesen.17,Ghosh.17}, we ignore the fluctuations on the tunneling.} In Fig.~\ref{fig:energy}, we plot the energy levels of $H_0$ as a function of $\epsilon_{q}$. As shown in the plot, in the region $\bar{\epsilon}_{q}\gg t_{p}$, the eigenstates for the ground state $|g\rangle$ and the first excited state $|e\rangle$ are approximately the states $|E\rangle$ and $|L\rangle$. In this work, we define the computational bases as the two-lowest eigenstates. In Fig.~\ref{fig:popu}, we further plot the population for the three eigenstates. The chosen operating point and the correspondingly defined qubit states not only benefit from the sweet spots, but also maximize the effective dipolar coupling between the qubit and the resonator (see Section.~\ref{sec:coupling} below). By introducing an external microwave-driven pulse on dipolar detuning around the sweet spot with $\Delta\epsilon_{d}=\epsilon(t)\cos(\omega_{0} t+\phi)$, when the frequency $\omega_{0}$ matches the qubit frequency $\omega_{ge}$, namely, on resonance, the total Hamiltonian can be reduced to an effective two-level structure in the interaction picture as (see Appendix~\ref{appx:B})
\begin{equation}
\begin{aligned}
H_{\rm{eff}}=\frac{\epsilon(t)}{2}(\cos\phi\ \sigma_{x}-\sin\phi\ \sigma_{y})
\end{aligned}
\label{eq:effective}
\end{equation}
\blue{Here, we emphasize again the logical bases are defined as $|0\rangle=|g\rangle$, $|1\rangle=|e\rangle$ within the operating regime $\bar{\epsilon}_{d}=t_{m}=0$ and $\bar{\epsilon}_{q}\gg t_{p}$, such that the Pauli matrix is $\sigma_{z}=|g\rangle\langle g|-| e\rangle\langle e|\approx|E\rangle\langle E|-| L\rangle\langle L|$.} In this way, arbitrary single-qubit gate can be implemented by using the two-lowest states as the computational basis. 
Note that, when $\bar{\epsilon}_{d}=t_{m}=0$, $H_0$ can also form a so-called CQ qubit in the bases $\{|E\rangle,|C\rangle\}$, leaving $|L\rangle$ as the leaked state \cite{Friesen.17}.

\section{Dipolar coupling to a resonator}\label{sec:coupling}

We first derive the dipole transition matrix element. Considering three quantum dots centering at $\bm{r}_{1}=-w\hat{x}$, $\bm{r}_{2}=0$, and $ \bm{r}_{3}=w\hat{x}$ respectively, the dipole operator for this triple-quantum-dot system is thus $\bm{d}=-e \sum_{i} \bm{r}_{i} n_{i} \equiv d \hat{x}$ ($i$=1,2,3), where $d=e w\left(n_{1}-n_{3}\right)$ and $n_{i}=|n_{i}\rangle\langle n_{i}|$. In the position bases, the dipole operator reads
\begin{equation}
\begin{aligned}
d=e w\left(|\mathit{100}\rangle\langle \mathit{100}|-| \mathit{001}\rangle\langle \mathit{001}|\right)=
e w\partial_{\epsilon_{d}} \mathcal{H}^{(0)}
\end{aligned}
\label{eq:dipole}
\end{equation}
which implies $n_{1}-n_{3}=\partial_{\epsilon_{d}}\mathcal{H}^{(0)}$. Introduce a small variation in the dipolar detuning $\epsilon_{d}=\bar{\epsilon}_{0}+\mathcal{F}+\delta \epsilon_{d}$, where $\bar{\epsilon}_{0}$ is the chosen operating point and $\mathcal{F}$ is the small variation. We assume $\mathcal{F}$ is with the order of $\delta \epsilon_{d}$. To simplify the discussion, in this section below we temporarily leave alone $\delta \epsilon_{d}$. Then we can expand the Hamiltonian $\mathcal{H}^{(0)}$ near the operating point as
\begin{equation}
\mathcal{H}\approx \mathcal{H}^{(0)}_{\epsilon_{d}=\bar{\epsilon}_{0}}+\left.\partial_{\epsilon_{d}} \mathcal{H}^{(0)}\right|_{\epsilon_{d}=\bar{\epsilon}_{0}} \mathcal{F},
\label{eq:Happro}
\end{equation}
where the first term of the right hand side of Eq.~(\ref{eq:Happro}) denotes the component of the Hamiltonian determined by $\bar{\epsilon}_{0}$, while the second term is proportional to the dipole operator in Eq.~(\ref{eq:dipole}), which we define as the dipole interaction Hamiltonian. In the computational bases, the reduced Hamiltonian for the small variation of the dipolar detuning can be rewritten as
\begin{equation}
\mathcal{H}=-\frac{ \omega_{ge}}{2} \sigma_{z}+\mathcal{F} \eta \sigma_{x}.
\label{eq:Happro2}
\end{equation}
Here, $\eta=\cos\theta$ and $\tan2\theta=2t_{p}/\bar{\epsilon}_{q}$, where we have assumed $\bar{\epsilon}_{0}=t_{m}=0$ (see Appendix.~\ref{appx:eigen}). Comparing Eqs.~(\ref{eq:Happro}) and (\ref{eq:Happro2}), one finds that
\begin{equation}
d_{ge}=\langle g|d| e\rangle=e w \eta=e w\cos\theta.
\label{eq:dipole2}
\end{equation}
Considering the operating region, $\bar{\epsilon}_{q}\gg t_{p}$, we have $\theta\sim0$, thus $d_{ge}$ is maximized. On the other hand, in Fig.~\ref{fig:popu}(c), we can see that in this region the component of the eigenstate $|f\rangle$ is dominated by $|C\rangle$. Namely, the components of the states $|E\rangle$ and $|L\rangle$ are close to zero. Therefore, $d_{ef}=\langle e|d| f\rangle\sim0$ and $d_{gf}=\langle g|d| f\rangle\sim0$, where $d_{mn}=\langle m|d| n\rangle$ is defined as the dipole transition matrix element. This means that, there is no transition between the second excited state $|f\rangle$ and other lower eigenstates induced by the small variation $\mathcal{F}$. In addition, one easily finds that $d_{gg}=d_{ee}=0$. Therefore, the TQD can be regarded as a well-defined two-level system spanned by $|E\rangle$ and $|L\rangle$ in this region.

Next, we determine the effective qubit-resonator coupling strength. We consider a TQD capacitively coupled to a transmission-line resonator with the lowest-energy mode as shown in Fig.~\ref{fig:quantumdot}(c), the geometry of which is similar to Ref.~\cite{Srinivasa.16}, and the coupling of the resonator field to the TQD is via  the variation in the dipolar detuning $\epsilon_{d}$. The quantized antinode voltage of the resonator is \cite{Childress.04}
\begin{equation}
\hat{V}=\sqrt{\frac{\hbar\omega_{r}}{LC_{0}}}(a+a^{\dagger})
\label{eq:qvolt}
\end{equation}
To make the derivation clear, we recover $\hbar$. Here, $a^{\dagger}$ ($a$) is the photon creation (annihilation) operator of the resonator, $\omega_{r}=\pi/(LZ_{0}C_{0})$ is the resonator frequency, $L$ denotes the length of the resonator, $C_{0}$ the capacitance per unit length. The characteristic impedance is $Z_{0}=\sqrt{L_{0} / C_{0}}$ with $L_{0}$ being the inductance per unit length. Further, the effective quantized voltage across the resonator is $\hat{V}_{\mathrm{eff}}=C_{c}\hat{V}/(C_{c}+C_{d})=\chi_{0}\hat{V}$, where $C_{c}$ is the total capacitance between the resonator and the TQD while $C_{d}$ denotes the capacitance between the TQD and the ground \cite{Srinivasa.16}. Therefore, the interaction between the TQD and the resonator is
\begin{eqnarray}\label{eq:Hint}
\mathcal{H}_{\mathrm{int}} &=& -\bm{d} \cdot \bm{E}=d \hat{V}_{\mathrm{eff}} / s \notag\\
 &=&\hbar g_{0}\left(n_{1}-n_{3}\right)\left(a+a^{\dagger}\right),
\end{eqnarray}
where
\begin{equation}
\begin{aligned}
g_{0}=\frac{e w \chi}{ s L C_{0}} \sqrt{\frac{\pi}{Z_{0} \hbar}}=\frac{e w \chi}{s} \omega_{r} \sqrt{\frac{Z_{0}}{\pi \hbar}}
\label{eq:g0}
\end{aligned}
\end{equation}
is the vacuum Rabi coupling strength and $s$ is the effective distance related to $\hat{V}_{\mathrm{eff}}$. On the other hand, comparing the dipole coupling Hamiltonian in Eqs.~(\ref{eq:Happro}) and (\ref{eq:Hint}), the small variation in the dipole detuning is equal to $\mathcal{F}=e w \hat{V}_{\mathrm{eff}} / s=\hbar g_{0}\left(a+a^{\dagger}\right)$. It is then clear that, the resonator controls the oscillation in the dipole detuning via its voltage and thus induces the transition between the qubit bases states. Moreover, substituting $\mathcal{F}$ into Eq.~(\ref{eq:Happro2}), the effective interaction in the computational bases can then be expressed as
\begin{equation}
\begin{aligned}
\tilde{\mathcal{H}}_{\mathrm{int}}= \hbar g \sigma_{x}\left(a+a^{\dagger}\right),
\label{eq:Heffint}
\end{aligned}
\end{equation}
where $g=g_{0}\eta=g_{0}\cos\theta$ is the effective coupling strength. As stated above, we have considered $\eta=\cos\theta=1$, i.e., $g=g_{0}$. To estimate the coupling strength, we consider $w=s/2$ and $\chi_{0}=0.28$ according to the data in Refs.~\cite{Blais.04,Childress.04,Srinivasa.16}. In addition, from Eq.~(\ref{eq:g0}), the coupling strength is proportional to $\sqrt{Z_{0}}$ and $\omega_{r}$. From the recent experiments~\cite{Stockklauser.17,Van.18, Wang.20}, where an array of high-impedance SQUID array is used to design the resonator, $Z_{0}$ can be as high as $1\ \mathrm{k}\Omega$. For a typical value of the resonator frequency $\omega_{r}/2\pi$ between $1.5$ and $6.5\ \rm{GHz}$, the coupling strength $g_{0}/2\pi\ $ is therefore in the range of [60, 250] MHz.


\section{Two-qubit entangling gates}\label{sec:twoqubitgate}
\blue{
Below, we present two approaches to construct the two-qubit entangling gates: (a) The two qubits are in resonance with each other, while they are detuned from the resonator. When working in the dispersive regime, i.e., $\Delta^{(k)}=\omega^{(k)}-\omega_{r}\gg g^{(k)}$, where $\Delta^{(k)}$ is defined as the qubit-resonator detuning, a dynamical iSWAP-type gate can be implemented. (b) Both qubits and the resonator are in resonance, namely, $\Delta^{(k)}=0$, in this way we can obtain a holonomic two-qubit entangling gate. }

\subsection{Dynamical iSWAP gate operated in the dispersive regime}

We now extend the discussion in Sec.~\ref{sec:coupling} that two separated charge qubits are coupled to the transmission-line resonator. The total Hamiltonian for this hybrid system consisting of two qubits and a resonator reads
\begin{equation}
	H_{\rm{tot}} = H_{\rm{res}}+\sum_{k=1}^2 \mathcal{H}_{0}^{(k)}+ \sum_{k=1}^2 \tilde{\mathcal{H}}_{\rm{int}}^{(k)}
	\label{eq:Htot1}
\end{equation}
where $\mathcal{H}_{0}^{(k)}$ is the Hamiltonian for the $k$th qubit (TQD) as described in Eq. (\ref{eq:Hpo}), $H_{\rm{res}}=\omega_{r}a^{\dagger}a$ is the Hamiltonian for the resonator, and $\tilde{\mathcal{H}}_{\rm{int}}^{(k)}$ represents the dipole interaction Hamiltonian between the $k$th qubit and the resonator as shown in Eq.~(\ref{eq:Heffint}). Transforming $H_{\rm{tot}}$ into the TQD eigenbasis, we have
\begin{equation}
	\begin{aligned}
		H_{\rm{tot}}=& \omega_{r} a^{\dagger} a+\sum_{k=1}^{2} \sum_{n=\{g,e,f\}} E_{n}^{(k)} \sigma_{n n}^{(k)} \\
		&+\sum_{k=1}^{2} \sum_{m,n=\{g,e,f\}} g^{(k)} d_{mn}^{(k)}\left(a+a^{\dagger}\right) \sigma_{ mn}^{(k)},
		\label{eq:Htoteigen}
	\end{aligned}
\end{equation}
where $\sigma_{mn}=|m\rangle\langle n|$.
As mentioned in Section~\ref{sec:coupling}, in the operating regime  $\bar{\epsilon}_{q}\gg t_{p}$, both qubits can be regarded as a well-defined two-level system, and $d_{gg}=d_{ee}=d_{ef}=d_{gf}\sim0$. Therefore, $H_{\rm{tot}}$ can be reduced to the so-called Tavis-Cummings form as \cite{Fink.09}
\begin{equation}
	H_{\rm{TC}}=\omega_{r}a^{\dagger}a-\sum_{k=1}^2 \left[\frac{\omega_{ge} ^{(k)}}{2} \sigma_{z}^{(k)} -g^{(k)} \sigma_{x}\left(a+a^{\dagger}\right)\right],
	\label{eq:Htot}
\end{equation}
where $\omega_{ge} ^{(k)}=E_{e}^{(k)}-E_{g}^{(k)}$ and $g^{(k)}$ represent the frequency and couping strength for the $k$th qubit, respectively. To simplify the discussion we set $\omega_{ge} ^{(k)}\equiv\omega^{(k)}$ hereafter.

For approach (a), we expand the discussion of Refs.~\cite{blais.07,Srinivasa.16} on the construction of the iSWAP gate when $\Delta^{(k)}\gg g^{(k)}$. The effective Hamiltonian for $H_{\rm{TC}}$ can be further simplified by using the Schrieffer-Wolff transformation \cite{blais.07}, which can eliminate the direct coupling between the qubit and the resonator:
\begin{equation}\\
	H_{d}=H_{d,0}+\frac{1}{2}\left[S, V\right],
	\label{eq:Hsw}
\end{equation}
where $H_{d,0}$ denotes the free Hamiltonian for the resonator and the two charge qubits:
\begin{equation}
	\begin{aligned}
		H_{d,0} &=\omega_{r}a^{\dagger}a- \sum_{k=1}^2 \frac{\omega ^{(k)}}{2} \sigma_{z}^{(k)},
	\end{aligned}
	\label{eq:Hdisper}
\end{equation}
$V$ the dipolar interaction Hamiltonian for individual qubit
\begin{equation}
	\begin{aligned}
		V=\sum_{k=1}^2g^{(k)} \sigma_{x}\left(a+a^{\dagger}\right),
	\end{aligned}
	\label{eq:Hdisperint}
\end{equation}
and $S$ the transformation operator
\begin{equation}
	S=\sum_{k=1}^2 \frac{g^{(k)}}{\Delta^{(k)}}(a^{\dagger}\sigma_-^{(k)}-\sigma_+^{(k)}a).
	\label{eq:S}
\end{equation}
Combining Eqs. (\ref{eq:Hsw}) and  (\ref{eq:S}), the resulting approximated Hamiltonian is
\begin{equation}
	\begin{aligned}
		H_{d}&\approx H_{d,0}+\sum_{k=1}^2 \frac{(g^{(k)})^{2}}{\Delta^{(k)}}(\sigma_-^{(k)}\sigma_+^{(k)}-\sigma_+^{(k)}\sigma_-^{(k)})a^{\dagger} a\\&-\frac{(g^{(k)})^{2}}{\Delta^{(k)}}\sigma_+^{(k)}\sigma_-^{(k)} -\chi(\sigma_+^{(1)}\sigma_-^{(2)}+\sigma_-^{(1)}\sigma_+^{(2)}),
	\end{aligned}
	\label{eq:Hdappro}
\end{equation}
where
$\chi= g^{(1)}g^{(2)}(\Delta^{(1)}+\Delta^{(2)})/[2\Delta^{(1)}\Delta^{(2)}]$. In the zero-photon subspace i.e., the computational subspace, $H_{d}$ can be further reduced to
\begin{equation}
	\begin{aligned}
		\tilde{H}_{d}&=\sum_{k=1}^2 \frac{\tilde{\omega}^{(k)}}{2} \sigma_{z}^{(k)}-\chi(\sigma_+^{(1)}\sigma_-^{(2)}+\sigma_-^{(1)}\sigma_+^{(2)}),
	\end{aligned}
	\label{eq:Hdtilde}
\end{equation}
where $\tilde{\omega}^{(k)}=-\omega^{(k)}+(g^{(k)})^{2}/\Delta^{(k)}$. Under the Schrieffer-Wolff transformation, the direct coupling between the qubit and the resonator has been safely eliminated and this approximation is correct to first order in $g^{(k)}/\Delta^{(k)} $. Further, we transform $\tilde{H}_{d}$ into a rotating frame via
\begin{equation}
	U_{d}=\exp\left[{-i\sum_{k=1}^2\frac{\tilde{\omega}^{(k)}}{2}\sigma_{z}^{(k)}t}\right],
	\label{eq:Urot}
\end{equation}
which leads to the Hamiltonian
\begin{equation}
	\begin{aligned}
		\tilde{H}_{d}&=U_{d}^{\dagger}\tilde{H}_{d}U_{d}-iU_{d}^{\dagger}\frac{\partial U_{d}}{\partial t}\\
		&=-\chi\left(\sigma_+^{(1)}\sigma_-^{(2)}+\sigma_-^{(1)}\sigma_+^{(2)}\right),
		\label{eq:Hd2}
	\end{aligned}
\end{equation}
where we have considered $\tilde{\omega}^{(1)}=\tilde{\omega}^{(2)}$. The evolution operator of the Hamiltonian $\tilde{H}_{d}$ is thus
\begin{equation}
	U_{\rm{ent}}'(t)=\begin{pmatrix}
		1 & 0  & 0 & 0 \\
		0 & \cos\chi t & i\sin\chi t & 0\\
		0 & i\sin\chi t  & \cos\chi t & 0 \\
		0 & 0  & 0 & 1
	\end{pmatrix}.
	\label{eq:Uiswap}
\end{equation}
When $\chi t=\pi/2$, $U_{\rm{ent}}'(\frac{\pi}{2\chi})$ is equivalent to an iSWAP gate.

\begin{figure}
	\includegraphics[width=1\columnwidth]{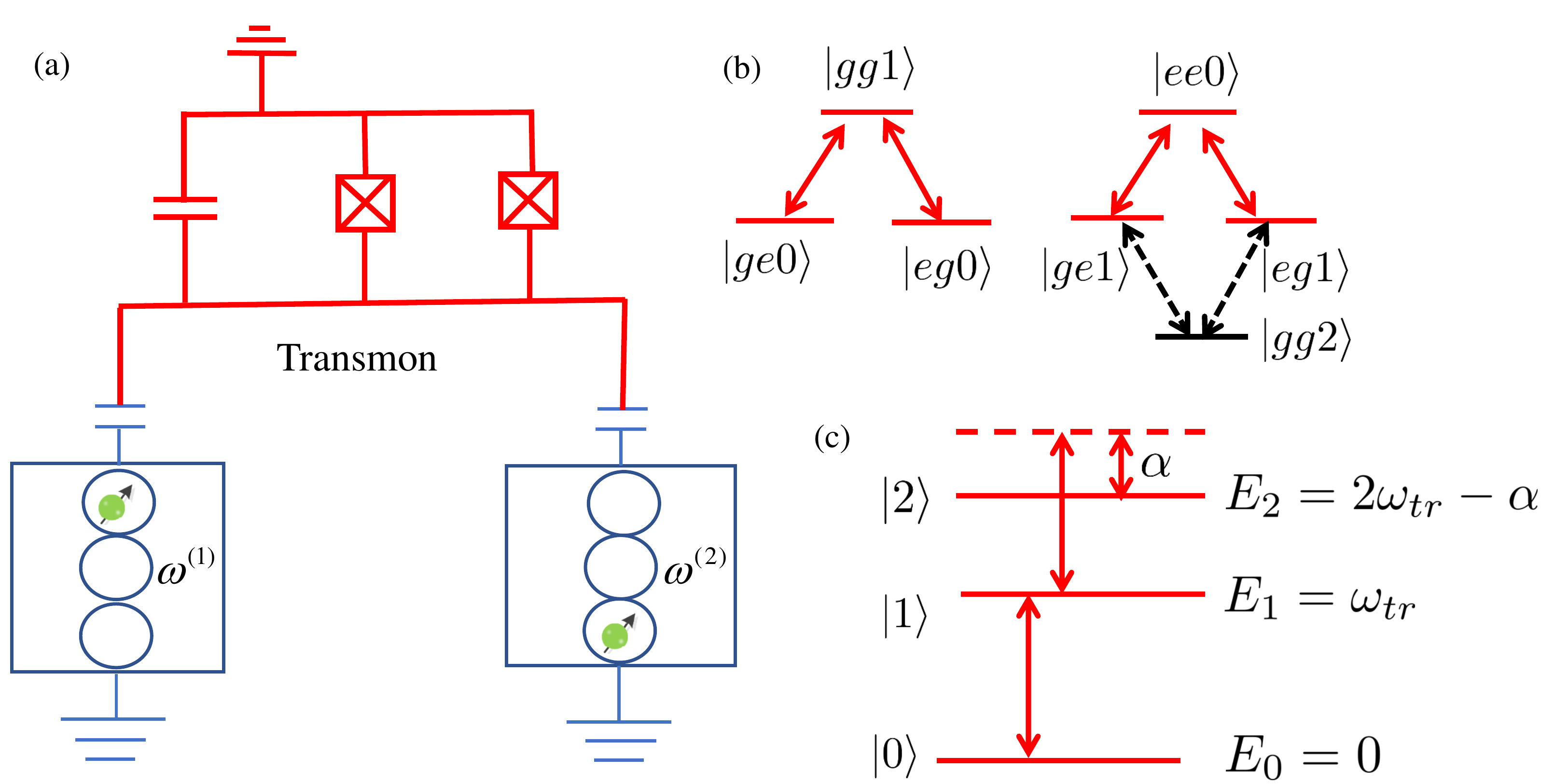}
	\caption{\blue{(a) The schematic of the coupling between the charge qubits and the transmon. (b) Possible transitions in the subspace $S_{1}$ and $S_{2}$, which form a three-level \bm{$\Lambda$} structure. In the subspace $S_{2}$, leakage to the state $|g,g,2\rangle$ occurs if the anharmonicity $\alpha$ is small. (c) Energy level for the transmon, which can be modeled as a resonator with nonlinear spectrum.}}
	\label{fig:transmon}
\end{figure}

\subsection{Holonomic gates operated in the resonant regime}

\blue{
As shown in Fig.~\ref{fig:transmon}(b), for the resonant case, the computational subspace is coupled to the leakage state, which is residing in the two-excited subspace (the detail is given below). To suppress this leakage, we introduce the dipolar coupling between the charge qubits and a superconducting transmon to employ its good anharmonic property \cite{scarlino.19,landig.19} (see Fig.~\ref{fig:transmon}(a)). The dipolar coupling model between the transmon and the charge qubits is similar to the case as shown in Section.~\ref{sec:coupling}. In this way, the Hamiltonian for the hybrid system including the transmon is slightly different from Eq.~\ref{eq:Htot1}:
\begin{equation}
	H_{\rm{tot}}' = \sum_{n=1}^{n}\left[n \omega_{tr}-\frac{n(n-1)}{2} \alpha\right]|n\rangle\langle n|    +\sum_{k=1}^2 \mathcal{H}_{0}^{(k)}+ \sum_{k=1}^2 \tilde{\mathcal{H}}_{\rm{int}}^{(k)}.
	\label{eq:Hr}
\end{equation}
Here, $n$ represents the $n$-th level of the transmon, which can be also regarded as the number in the photon for the resonator. While $\omega_{tr}$ and $\alpha$ refer to the intrinsic frequency and anharmonicity for the transmon, respectively. From Eq.~\ref{eq:Hr}, the transmon can be modeled as a resonator with nonlinear spectrum
\begin{equation}
	H_{\rm{tot}}'' = \omega_{tr}a^{\dagger}a-\sum_{n=1}^{n}\left[\frac{n(n-1)}{2} \alpha\right]|n\rangle\langle n|    +\sum_{k=1}^2 \mathcal{H}_{0}^{(k)}+ \sum_{k=1}^2 \tilde{\mathcal{H}}_{\rm{int}}^{(k)}.
	\label{eq:Hr2}
\end{equation}
For the resonant case, i.e., $\omega^{(k)}=\omega_{tr}$, in the interaction picture defined by $U_{tr}=\exp[-i H_{tr}^{0}t ]$ where
\begin{equation}
H_{tr}^{0}=\omega_{tr}a^{\dagger}a-\sum_{n=1}^{n}\left[\frac{n(n-1)}{2} \alpha\right]|n\rangle\langle n| +\sum_{k=1}^2 \mathcal{H}_{0}^{(k)},
\label{eq:Ur}
\end{equation}
we have
\begin{equation}
\begin{aligned}
H_{tr}\simeq&\sum_{k=1}^2g^{(k)}\left(a^{\dagger}\sigma_-^{(k)}+\mathrm{H.c.}\right).
\label{eq:Htotr}
\end{aligned}
\end{equation}
Note that here we have ignored the fast-oscillating terms $g^{(k)}\exp[\pm i \alpha t]$ related to the transmon state $|2\rangle$ by assuming $g^{(k)}\ll \alpha$ (the small anharmonicity effect will be discussed later). Also, we have ignored the energy-nonconserving terms $a\sigma_-^{(k)}$ and $a^{\dagger}\sigma_+^{(k)}$, due to $ \omega_{r}, \omega_{r}-\alpha\gg g$. Further, the higher level of the transmon will be strongly suppressed. Ideally, we can design the holonomic gate considering the photon subspace $n=0,1$.} Because the total number of excitations is conserved, we can further rewrite $H_{tr}$ in a block-diagonal form. In the single- and two-excited subspaces, we have $S_{1}=\operatorname{span}\{|e,g,0\rangle,|g,e,0\rangle,|g,g,1\rangle\}$ and $S_{2}=\operatorname{span}\{|e,e,0\rangle,|e,g,1\rangle,|g,e,1\rangle\}$, where the qubit states from the left to the right denote qubit 1 and 2, respectively. The corresponding Hamiltonian in these two subspaces have the similar forms as
\begin{equation}\label{eq:Hrspace1}
\mathcal{H}_{tr,1}=\left|g,g,1\right\rangle (g^{(1)}\left\langle e,g,0\right|+g^{(2)} \left\langle g,e,0\right|)+\mathrm{H.c.}
\end{equation}
and
\begin{equation}\label{eq:Hrspace2}
\mathcal{H}_{tr,2}=\left|e,e,0\right\rangle (g^{(1)}\left\langle g,e,1\right|+g^{(2)} \left\langle e,g,1\right|)+\mathrm{H.c.}
\end{equation}
As shown in Fig.~\ref{fig:transmon}(b), $\mathcal{H}_{tr,1}$ can form a three-level \bm{$\Lambda$} structure \cite{hong.18, Egger.19,li.20,Zhang.20} with transitions between $|g,g,1\rangle\leftrightarrow\left|e,g,0\right\rangle$ and $|g,g,1\rangle\leftrightarrow|g,e,0\rangle$. Similarly, $\mathcal{H}_{tr,2}$ introduces such transition between $\left|e,e,0\right\rangle\leftrightarrow|e,g,1\rangle$ and \blue{$|e,e,0\rangle\leftrightarrow|g,e,1\rangle$.} \blue{In fact, $\mathcal{H}_{tr}$ in the two-excited subspace can also induce transition between $|g,g,2\rangle\leftrightarrow|g,e,1\rangle$ and $|g,g,2\rangle\leftrightarrow|e,g,1\rangle$ for the small anharmonicity (see Eq.~\ref{fig:transmon}(b)).} The remaining two subspaces are $S_{3}=\operatorname{span}\{\left| {g,g,0} \right\rangle \}$ and $S_{4}=\operatorname{span}\{\left| {e,e,1} \right\rangle \}$, with $\mathcal{H}_{tr,3}=\mathcal{H}_{tr,4}=0$ due to $\Delta^{(k)}=0$.

To make the derivation clear, below we follow Ref.~\cite{Zhang.20} to expand the discussion on how to implement the holonomic operation using $\mathcal{H}_{tr,1}$. The case for $\mathcal{H}_{tr,2}$ can be understood in the same way since they have the similar Hamiltonian structure. $\mathcal{H}_{tr,1}$ can also be expressed using a bright-dark representation
\begin{equation}
\begin{aligned}
\mathcal{H}_{tr,1}=\Omega \left|g,g,1\right\rangle \langle b|+\mathrm{H.c.}
\end{aligned}
\end{equation}
 where
\begin{eqnarray}
|b\rangle&=&\sin\frac{\varphi}{2}|e,g,0\rangle -\cos\frac{\varphi}{2}|g,e,0\rangle\notag
\label{eq:dressedb}
\end{eqnarray}
is the bright state, while
\begin{eqnarray}
|d\rangle&=&\cos\frac{\varphi}{2}|e,g,0\rangle +\sin\frac{\varphi}{2}|g,e,0\rangle
\label{eq:dressedd}
\end{eqnarray}
is the dark state.
$\Omega=\sqrt{\left(g^{(1)}\right)^{2}+\left(g^{(2)}\right)^{2}}$ and $\tan\varphi/2=-g^{(1)}/g^{(2)}$. In this representation, the dark state $|d\rangle$ has been dropped out of the dynamics. Therefore, $\mathcal{H}_{tr,1}$ can be regarded as the transitions between the bright state $|b\rangle$ and state $|g,g,1\rangle$. Thus, the evolution operator with respect to $\mathcal{H}_{tr,1}$ is
\begin{eqnarray}\label{eq:Ubd}
U_{tr,1}(t)&=&\exp\left(-i\int_{0}^{t} \mathcal{H}_{tr,1}dt^{\prime}\right)\notag\\
&=&\cos\delta(t) (|g,g,1\rangle\langle g,g,1|+|b\rangle\langle b|)\\
&&-i\sin\delta(t)(|g,g,1\rangle\langle b|+|b\rangle\langle g,g,1|)+|d\rangle\langle d|,\notag
\end{eqnarray}
where $\delta(t)=\int_{0}^{t} \Omega \mathrm{d} t^{\prime}$. According to Eq. (\ref{eq:Ubd}), when the cyclic condition is met, i.e., $\delta(T)=\pi$, the evolution operator in the subspace $S_{1}$ is
\begin{equation}
U_{tr,1}(T)=\begin{aligned}
\left( {\begin{array}{*{20}{c}}
	\cos\varphi&\sin\varphi&0\\
	\sin\varphi&-\cos\varphi&0\\
	0&0&-1\\
	\end{array}} \right)
\end{aligned}
\label{eq:Ur3T}
\end{equation}
As we can see that after the cyclic evolution, the operator matrix is block-diagonalized. Therefore, the excited state of the resonator  $|g,g,1\rangle$, cannot affect the qubit subspace spanned by $\{|e,g,0\rangle,|g,e,0\rangle\}$. For an arbitrary state initialized in the subspace spanned by $S_{i}=\operatorname{span}\left\{\left|\psi_{1}(0)\right\rangle,\left|\psi_{2}(0)\right\rangle\right\}$, where
\begin{equation}
\begin{aligned}
\left|\psi_{1}(0)\right\rangle &=\alpha|e,g,0\rangle+\beta|g,e,0\rangle, \\
\left|\psi_{2}(0)\right\rangle &=\beta^{*}|e,g,0\rangle-\alpha^{*}|g,e,0\rangle,
\end{aligned}
\label{eq:Sint}
\end{equation}
the corresponding final states under the action of $\mathcal{H}_{tr,1}$ satisfy the parallel-transport condition for the holonomic gate \cite{Erik.12,Sjoqvist.15}, i.e, $\left\langle\psi_{i}(t)\left|\mathcal{H}_{r,1}(t)\right| \psi_{j}(t)\right\rangle=\left\langle\psi_{i}(0)\left|U_{tr,1}^{\dagger}(t)\left|\mathcal{H}_{tr,1}(t)\right| U_{tr,1}(t)\right|\psi_{j}(0)\right\rangle=0$. Here, $\alpha$, $\beta$ $\in \mathbb{C}$ and $|\alpha|^{2}+|\beta|^{2}=1$. Therefore, $U_{tr,1}(T)$ represents a holonomic operation in the qubit subspace $\{|e,g,0\rangle,|g,e,0\rangle\}$. Then, in the complete computational subspace (also the zero-photon subspace) $\left\{\left|g,g,0\right\rangle, \left|g,e,0\right\rangle,\left|e,g,0\right\rangle,\left|e,e,0\right\rangle\right\}$, we have the holonomic two-qubit gate as
\begin{equation}
U_{\rm{ent}}(\varphi)=\begin{pmatrix}
1 & 0  & 0 & 0 \\
0 & \cos\varphi  & \sin\varphi & 0\\
0 & \sin\varphi & -\cos\varphi & 0 \\
0 & 0  & 0 & -1
\end{pmatrix}.
\label{eq:Uent}
\end{equation}
Note that the negative sign in the bottom right is owing to the evolution in the two-excited subspace $\mathcal{H}_{r,2}$ \cite{Zhou.18}. As demonstrated in Ref.~\cite{Zhang.20}, $U_{\rm{ent}}(\pi/2)$, which corresponds to $g^{(1)}=g^{(2)}=g$, denotes an iSWAP-type two-qubit entangling gate.

\begin{figure}
	\includegraphics[width=1\columnwidth]{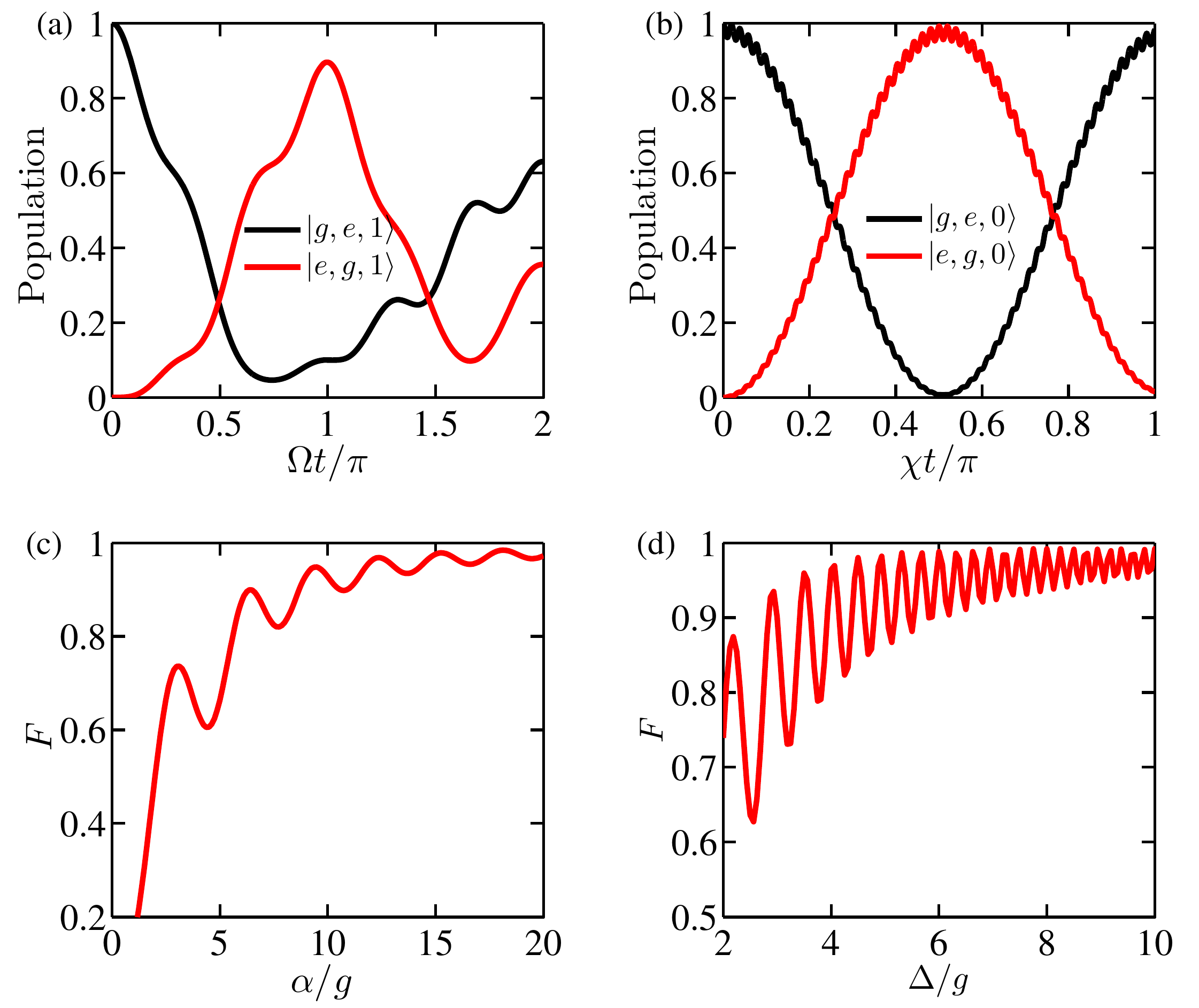}
	\caption{Fidelity and state population of the holonomic entangling gate $U_{\rm{ent}}(\varphi=\pi/2)$ (left column) and the iSWAP gate $U_{\rm{ent}}'(t=\frac{\pi}{2\chi})$ (right column). The state population for the holonomic and iSWAP entangling gate are shown in (a) and (b), respectively. \blue{ (c) Fidelity of holonomic entangling gate as a function of $\alpha/g$. (d) Fidelity of the iSWAP gate as a function of the ratio $\Delta/g$. The common parameters for all the panels used in the simulation: $g/2\pi=66\ \rm{MHz}$, $\Gamma_{\varphi}^{(1)}/2\pi=\Gamma_{\varphi}^{(2)}/2\pi=2.7\ \rm{MHz}$, $\Gamma_{ge}=\Gamma_{ef}=\Gamma_{gf}=0$ \cite{scarlino.19}. The other parameters: (a) $\Delta=0$, $\Gamma_{a,tr}/2\pi=4\ \rm{KHz}$ \cite{Zi.16,Tao.18}, $\Gamma_{\varphi,tr}/2\pi=0.8\ \rm{MHz}$ \cite{scarlino.19}. (b) $\Delta/g=10$ and $\Gamma_{a,r}/2\pi=0.028\ \rm{MHz}$ \cite{Samkharadze.16}.}}
	\label{fig:twoqubit}
\end{figure}

\subsection{Gate fidelity} 
To simulate the gate fidelity for the entangling gates, we consider using the master equation as \cite{blais.07}
\begin{eqnarray}
\dot{\rho}=-i\left[H_{\rm{tot}}, \rho\right]+\mathcal{L}_{1} \rho+\mathcal{L}_{\varphi} \rho+\mathcal{L}_{a} \rho,
\label{eq:master2}
\end{eqnarray}
where
\begin{equation}
\begin{aligned}
\mathcal{L}_{1} \rho&=\sum_{k=1,2}\Gamma_{ef}^{(k)} \mathcal{D}[|e\rangle\langle f|]+\Gamma_{gf}^{(k)} \mathcal{D}[|g\rangle\langle f|]+\Gamma_{ge}^{(k)} \mathcal{D}[|g\rangle\langle e|], \\
\mathcal{L}_{\varphi} \rho&=\sum_{k=1,2}\frac{1}{2}\Gamma_{\varphi}^{(k)} \mathcal{D}[| e\rangle\langle e|-| g\rangle\langle g|], \\
\mathcal{L}_{a} \rho&=\Gamma_{a} \mathcal{D}[a],
\end{aligned}
\label{eq:linblad2}
\end{equation}
and
\begin{equation}
\begin{aligned}\mathcal{D}[\hat{L}]=\left(2 L \rho L^{\dagger}-L^{\dagger} L \rho-\rho L^{\dagger} L\right) / 2.
\label{eq:D}
\end{aligned}
\end{equation}
Here, $\mathcal{L}_{1} \rho$ and $\mathcal{L}_{\varphi}\rho$ denote the possible relaxation and dephasing baths for each qubit, while $\mathcal{L}_{a}\rho$ the decay of the resonator. \blue{From the recent experiment in Ref.~\cite{scarlino.19}, we consider the coupling strength  $g^{(1)}=g^{(2)}=g=2\pi\times66\ \rm{MHz}$ (corresponding to $\omega_{r}/2\pi\sim1.7\ \rm{GHz}$), and the dephasing rate $\Gamma_{\varphi}^{(1)}/2\pi=\Gamma_{\varphi}^{(2)}/2\pi=2.7\ \rm{MHz}$.} While we set the qubit relaxation rate to be zero, i.e., $\Gamma_{ge}=\Gamma_{ef}=\Gamma_{gf}=0$. Because in the experiment, the linewidths of the qubits can be directly measured, which includes the relaxation effect of the qubits. In addition, according to the work \cite{Samkharadze.16}, \blue{the decay of the resonator can be as low as $\Gamma_{a,r}/2\pi=0.028\ \rm{MHz}$, which corresponds to the quality factor of $10^{5}$. For the iSWAP gate, we consider the initial state of the coupled system as $|g,e,0\rangle$. Ideally, the final state is expected to be $|e,g,0\rangle$ without decoherence effect. In Fig.~\ref{fig:twoqubit}(d), we plot the fidelity of the iSWAP gate as a function of $\Delta/g$. The fidelity is defined as $F=\rm{Tr}\left[\rho_{\rm{id}}. \rho_{\rm{re}}\right]$, where $\rho_{\rm{id}}$ and $ \rho_{\rm{re}}$ denote the ideal and realistic density matrix, respectively. Here, we consider the qubit-resonator detuning $\Delta^{(1)}=\Delta^{(2)}=\Delta$. We find that, the fidelity is increasing as the detuning becomes large. When $\Delta/g=10$, the related fidelity surpass 99.2\%. The corresponding population is shown in Fig.~\ref{fig:twoqubit}(b).
As shown in Fig.~\ref{fig:twoqubit}(c), we plot the fidelity for the holonomic gate as a function of $\alpha/g$. In the simulation, we take a normal decay rate of the transmon as $\Gamma_{a,tr}/2\pi=4\ \rm{KHz}$ \cite{Zi.16,Tao.18}, while the dephasing rate for the transmon is $\Gamma_{\varphi,tr}/2\pi=0.8\ \rm{MHz}$ \cite{scarlino.19} with $\mathcal{L}_{\varphi,tr} \rho=\frac{1}{2}\Gamma_{tr} \mathcal{D}[| 0\rangle\langle 0|-| 1\rangle\langle 1|]$. The other parameters are similar to the case for the iSWAP gate for fair comparison. It is clear that the performance of the holonomic gate sensitively depends on the value of the anharmonicity. When the anharmonicity is zero the fidelity can be as low as 0.4, due to sever leakage to the state $|g,g,2\rangle$. Obviously, the fidelity gradually increase as the anharmonicity is increasing. When the anharmonicity is large enough with $\alpha/g\geq10$, the fidelity can reach about 98\%. However, in the experiment, the large anharmonicity would cause unwanted charge noise for the transmon. Normally, the anharmonicity is with the range of $\alpha/2\pi=[200, 400]$ MHz \cite{Zhao.20}. In Fig.~\ref{fig:twoqubit}(a), we show the population for the holonomic gate considering $\alpha/2\pi\simeq 400$ MHz (corresponding to $\alpha/g\simeq6.2$), the related fidelity is about 90\%. Note that since the leakage to the state $|g,g,2\rangle$ only affect the subspace $S_{2}$, here we consider the initial state to be $|g,e,1\rangle$ rather than $|g,e,0\rangle$.}


\section{Conclusion}

\blue{
We have proposed the implementation of a new type of charge qubit formed by an electron confined in a triple-quantum-dot system, which can work at the dipolar and quadrupolar detuning sweet spots. Particularly, we propose how to couple two separated charge qubits in a TQD via the superconducting resonator, where two types of entangling gates, i.e., the iSWAP and the holonomic gates are implemented. We find that the fidelity for the iSWAP gate can surpass 99\% considering the noise level in experiments. While the fidelity for the Holonomic gate can reach 98\%, if the anharmonicity in the resonator is large enough. To conclude, our proposal might offer an alternative way to implement high-fidelity quantum control for charge qubits based on semiconductor quantum dot.}

\section*{ACKNOWLEDGMENTS}
\blue{We thank Tao Chen for useful discussion.} This work was supported by the Key-Area Research and Development Program of Guang Dong Province  (Grant No. 2018B030326001), the National Natural Science Foundation of China (Grant Nos. 11905065, 11874156, 11874312), the Research Grants Council of Hong Kong (No. CityU 11303617), the Guang Dong Innovative and Entrepreneurial Research Team Program (No. 2016ZT06D348), \blue{and the Guangxi Science Foundation (Grant No. AD22035186).}

\appendix

\section{Eigenstates for TQD}\label{appx:eigen}
When $\bar{\epsilon}_{d}=t_{m}=0$, the eigenvalues for $H_0$ is
\begin{equation}
\begin{aligned}
E_{g} &=\left(\bar{\epsilon}_{q}-\sqrt{4t_{p}^{2}+\bar{\epsilon}_{q}^{2}}\right)/2,\\
E_{e} &=0, \\
E_{f} &=\left(\bar{\epsilon}_{q}+\sqrt{4t_{p}^{2}+\bar{\epsilon}_{q}^{2}}\right)/2,
\end{aligned}
\label{eq:eigenvalue}
\end{equation}
which corresponds to the eigenstates as
\begin{equation}
\begin{aligned}
|g\rangle &=\cos\theta|E\rangle-\sin\theta|C\rangle,\\
|e\rangle &=|L\rangle, \\
|f\rangle &=\sin\theta|E\rangle+\cos\theta|C\rangle,
\end{aligned}
\label{eq:eigenstate}
\end{equation}
where $|g\rangle$ is the ground state while $|e\rangle$ and $|f\rangle$ are the first and second excited state, respectively. The energy difference between the related states are
\begin{equation}
\begin{aligned}
\omega_{ge}=E_{e}-E_{g} &=\left(\sqrt{4t_{p}^{2}+\bar{\epsilon}_{q}^{2}}-\bar{\epsilon}_{q}\right)/2,\\
\omega_{gf}=E_{f}-E_{g} &=\sqrt{4t_{p}^{2}+\bar{\epsilon}_{q}^{2}},\\
\omega_{ef}=E_{f}-E_{e} &=E_{f}.
\end{aligned}
\label{eq:eigeneg}
\end{equation}
Note that, here we have considered $t_{12}=t_{23}$ and thus $\tan2\theta=2t_{p}/\bar{\epsilon}_{q}$. In the regime where $t_{p}\ll\bar{\epsilon}_{q}$, $\sin\theta\sim0$ and $\cos\theta\sim1$, we have $| g\rangle\sim |E\rangle$ and $| f\rangle\sim |C\rangle$.

\section{Effective Hamiltonian for the TQD}\label{appx:B}

The Hamiltonian including the microwave-driven pulse on dipolar detuning has three parts
\begin{equation}
\begin{aligned}
H_{\rm{eff}}&=H_0+H'+H_{\rm{m}},
\end{aligned}
\label{eq:effecth0}
\end{equation}
where
\begin{equation}
\begin{aligned}
H_0&=E_{g}|g\rangle \langle g|+E_{e}|e\rangle \langle e|+E_{f}|f\rangle \langle f|,\\
H'&=\delta\epsilon_{q}|C\rangle \langle C|+\delta\epsilon_{d}(|E\rangle \langle L|+|L\rangle \langle E|),\\
H_{\rm{m}}&=\epsilon(t)\cos(\omega_{0} t+\phi)(|E\rangle \langle L|+|L\rangle \langle E|).
\end{aligned}
\label{eq:effecth}
\end{equation}
We assume both the charge noises and the microwave-driven pulse $\epsilon(t)$ are much smaller compared to $t_{p}$ and $\bar{\epsilon}_{q}$. Therefore, $H'$ and $H_{\rm{m}}$ are regarded as the perturbation. In the interaction picture defined by $U_{0}=e^{-i H_{0}t} $ we have
\begin{widetext}
\begin{equation}
\begin{aligned}
H_{\rm{eff}}&=U_{0}^{\dagger}(H'+H_{\rm{m}})U_{0}\\
&\approx\left(\begin{array}{ccc}
\delta\epsilon_{q} \sin^{2}\theta & \cos \theta\left(\frac{\epsilon(t)}{2} e^{i \phi}+\delta\epsilon_{d}\right) & 0 \\
\cos\theta\left(\frac{\epsilon(t)}{2} e^{-i \phi}+\delta\epsilon_{d} \right) & 0 & 0 \\
0 & 0 & \delta\epsilon_{q} \cos^{2}\theta
\end{array}\right)
\label{eq:Hinth}
\end{aligned}
\end{equation}
\end{widetext}
in the eigenstates basis. Here, the counter rotating terms have been neglected in the assumption of $|\omega_{ef}-\omega_{0}|\gg \epsilon(t)$, $|\omega_{ge}+\omega_{0}|\gg \epsilon(t)$, $|\omega_{ge}+\omega_{0}|\gg \delta\epsilon_{d}, \delta\epsilon_{q}$, and $|\omega_{ge}|=\omega_{0}$.
Further, in the absence of noise and $\epsilon_{q}\gg t_{p}$ ($\cos\theta\sim1$) we have
\begin{equation}
\begin{aligned}
H_{\rm{eff}}&=\frac{\epsilon(t)}{2}(\cos\phi\ \sigma_{x}-\sin\phi\ \sigma_{y}).
\label{eq:Hinth2}
\end{aligned}
\end{equation}


\begin{thebibliography}{39}%
\makeatletter
\providecommand \@ifxundefined [1]{%
 \@ifx{#1\undefined}
}%
\providecommand \@ifnum [1]{%
 \ifnum #1\expandafter \@firstoftwo
 \else \expandafter \@secondoftwo
 \fi
}%
\providecommand \@ifx [1]{%
 \ifx #1\expandafter \@firstoftwo
 \else \expandafter \@secondoftwo
 \fi
}%
\providecommand \natexlab [1]{#1}%
\providecommand \enquote  [1]{``#1''}%
\providecommand \bibnamefont  [1]{#1}%
\providecommand \bibfnamefont [1]{#1}%
\providecommand \citenamefont [1]{#1}%
\providecommand \href@noop [0]{\@secondoftwo}%
\providecommand \href [0]{\begingroup \@sanitize@url \@href}%
\providecommand \@href[1]{\@@startlink{#1}\@@href}%
\providecommand \@@href[1]{\endgroup#1\@@endlink}%
\providecommand \@sanitize@url [0]{\catcode `\\12\catcode `\$12\catcode
  `\&12\catcode `\#12\catcode `\^12\catcode `\_12\catcode `\%12\relax}%
\providecommand \@@startlink[1]{}%
\providecommand \@@endlink[0]{}%
\providecommand \url  [0]{\begingroup\@sanitize@url \@url }%
\providecommand \@url [1]{\endgroup\@href {#1}{\urlprefix }}%
\providecommand \urlprefix  [0]{URL }%
\providecommand \Eprint [0]{\href }%
\providecommand \doibase [0]{http://dx.doi.org/}%
\providecommand \selectlanguage [0]{\@gobble}%
\providecommand \bibinfo  [0]{\@secondoftwo}%
\providecommand \bibfield  [0]{\@secondoftwo}%
\providecommand \translation [1]{[#1]}%
\providecommand \BibitemOpen [0]{}%
\providecommand \bibitemStop [0]{}%
\providecommand \bibitemNoStop [0]{.\EOS\space}%
\providecommand \EOS [0]{\spacefactor3000\relax}%
\providecommand \BibitemShut  [1]{\csname bibitem#1\endcsname}%
\let\auto@bib@innerbib\@empty
\bibitem [{\citenamefont {Shinkai}\ \emph {et~al.}(2009)\citenamefont
  {Shinkai}, \citenamefont {Hayashi}, \citenamefont {Ota},\ and\ \citenamefont
  {Fujisawa}}]{Shinkai.09}%
  \BibitemOpen
  \bibfield  {author} {\bibinfo {author} {\bibfnamefont {G.}~\bibnamefont
  {Shinkai}}, \bibinfo {author} {\bibfnamefont {T.}~\bibnamefont {Hayashi}},
  \bibinfo {author} {\bibfnamefont {T.}~\bibnamefont {Ota}}, \ and\ \bibinfo
  {author} {\bibfnamefont {T.}~\bibnamefont {Fujisawa}},\ }\href
  {https://link.aps.org/doi/10.1103/PhysRevLett.103.056802} {\bibfield
  {journal} {\bibinfo  {journal} {Phys. Rev. Lett.}\ }\textbf {\bibinfo
  {volume} {103}},\ \bibinfo {pages} {056802} (\bibinfo {year}
  {2009})}\BibitemShut {NoStop}%
\bibitem [{\citenamefont {Petersson}\ \emph {et~al.}(2010)\citenamefont
  {Petersson}, \citenamefont {Petta}, \citenamefont {Lu},\ and\ \citenamefont
  {Gossard}}]{Petersson.10}%
  \BibitemOpen
  \bibfield  {author} {\bibinfo {author} {\bibfnamefont {K.~D.}\ \bibnamefont
  {Petersson}}, \bibinfo {author} {\bibfnamefont {J.~R.}\ \bibnamefont
  {Petta}}, \bibinfo {author} {\bibfnamefont {H.}~\bibnamefont {Lu}}, \ and\
  \bibinfo {author} {\bibfnamefont {A.~C.}\ \bibnamefont {Gossard}},\ }\href
  {https://dx.doi.org/10.1103/PhysRevLett.105.246804} {\bibfield  {journal}
  {\bibinfo  {journal} {Phys. Rev. Lett.}\ }\textbf {\bibinfo {volume} {105}},\
  \bibinfo {pages} {246804} (\bibinfo {year} {2010})}\BibitemShut {NoStop}%
\bibitem [{\citenamefont {Cao}\ \emph {et~al.}(2013)\citenamefont {Cao},
  \citenamefont {Li}, \citenamefont {Tu}, \citenamefont {Wang}, \citenamefont
  {Zhou}, \citenamefont {Xiao}, \citenamefont {Guo}, \citenamefont {Jiang},\
  and\ \citenamefont {Guo}}]{Cao.13}%
  \BibitemOpen
  \bibfield  {author} {\bibinfo {author} {\bibfnamefont {G.}~\bibnamefont
  {Cao}}, \bibinfo {author} {\bibfnamefont {H.-O.}\ \bibnamefont {Li}},
  \bibinfo {author} {\bibfnamefont {T.}~\bibnamefont {Tu}}, \bibinfo {author}
  {\bibfnamefont {L.}~\bibnamefont {Wang}}, \bibinfo {author} {\bibfnamefont
  {C.}~\bibnamefont {Zhou}}, \bibinfo {author} {\bibfnamefont {M.}~\bibnamefont
  {Xiao}}, \bibinfo {author} {\bibfnamefont {G.-C.}\ \bibnamefont {Guo}},
  \bibinfo {author} {\bibfnamefont {H.-W.}\ \bibnamefont {Jiang}}, \ and\
  \bibinfo {author} {\bibfnamefont {G.-P.}\ \bibnamefont {Guo}},\ }\href
  {\doibase 10.1038/ncomms2412} {\bibfield  {journal} {\bibinfo  {journal}
  {Nat. Commun.}\ }\textbf {\bibinfo {volume} {4}},\ \bibinfo {pages} {1401}
  (\bibinfo {year} {2013})}\BibitemShut {NoStop}%
\bibitem [{\citenamefont {Li}\ \emph {et~al.}(2015)\citenamefont {Li},
  \citenamefont {Cao}, \citenamefont {Yu}, \citenamefont {Xiao}, \citenamefont
  {Guo}, \citenamefont {Jiang},\ and\ \citenamefont {Guo}}]{Li.15}%
  \BibitemOpen
  \bibfield  {author} {\bibinfo {author} {\bibfnamefont {H.-O.}\ \bibnamefont
  {Li}}, \bibinfo {author} {\bibfnamefont {G.}~\bibnamefont {Cao}}, \bibinfo
  {author} {\bibfnamefont {G.-D.}\ \bibnamefont {Yu}}, \bibinfo {author}
  {\bibfnamefont {M.}~\bibnamefont {Xiao}}, \bibinfo {author} {\bibfnamefont
  {G.-C.}\ \bibnamefont {Guo}}, \bibinfo {author} {\bibfnamefont {H.-W.}\
  \bibnamefont {Jiang}}, \ and\ \bibinfo {author} {\bibfnamefont {G.-P.}\
  \bibnamefont {Guo}},\ }\href {\doibase 10.1038/ncomms8681} {\bibfield
  {journal} {\bibinfo  {journal} {Nat. Commun.}\ }\textbf {\bibinfo {volume}
  {6}},\ \bibinfo {pages} {7681} (\bibinfo {year} {2015})}\BibitemShut
  {NoStop}%
\bibitem [{\citenamefont {Kim}\ \emph {et~al.}(2015)\citenamefont {Kim},
  \citenamefont {Ward}, \citenamefont {Simmons}, \citenamefont {Gamble},
  \citenamefont {Blume-Kohout}, \citenamefont {Nielsen}, \citenamefont
  {Savage}, \citenamefont {Lagally}, \citenamefont {Friesen}, \citenamefont
  {Coppersmith} \emph {et~al.}}]{Kim.15}%
  \BibitemOpen
  \bibfield  {author} {\bibinfo {author} {\bibfnamefont {D.}~\bibnamefont
  {Kim}}, \bibinfo {author} {\bibfnamefont {D.}~\bibnamefont {Ward}}, \bibinfo
  {author} {\bibfnamefont {C.}~\bibnamefont {Simmons}}, \bibinfo {author}
  {\bibfnamefont {J.~K.}\ \bibnamefont {Gamble}}, \bibinfo {author}
  {\bibfnamefont {R.}~\bibnamefont {Blume-Kohout}}, \bibinfo {author}
  {\bibfnamefont {E.}~\bibnamefont {Nielsen}}, \bibinfo {author} {\bibfnamefont
  {D.}~\bibnamefont {Savage}}, \bibinfo {author} {\bibfnamefont
  {M.}~\bibnamefont {Lagally}}, \bibinfo {author} {\bibfnamefont
  {M.}~\bibnamefont {Friesen}}, \bibinfo {author} {\bibfnamefont
  {S.}~\bibnamefont {Coppersmith}},  \emph {et~al.},\ }\href {\doibase
  10.1038/NNANO.2014.336} {\bibfield  {journal} {\bibinfo  {journal} {Nat.
  Nanotechnol.}\ }\textbf {\bibinfo {volume} {10}},\ \bibinfo {pages} {243}
  (\bibinfo {year} {2015})}\BibitemShut {NoStop}%
\bibitem [{\citenamefont {Ward}\ \emph {et~al.}(2016)\citenamefont {Ward},
  \citenamefont {Kim}, \citenamefont {Savage}, \citenamefont {Lagally},
  \citenamefont {Foote}, \citenamefont {Friesen}, \citenamefont {Coppersmith},\
  and\ \citenamefont {Eriksson}}]{Ward.16}%
  \BibitemOpen
  \bibfield  {author} {\bibinfo {author} {\bibfnamefont {D.~R.}\ \bibnamefont
  {Ward}}, \bibinfo {author} {\bibfnamefont {D.}~\bibnamefont {Kim}}, \bibinfo
  {author} {\bibfnamefont {D.~E.}\ \bibnamefont {Savage}}, \bibinfo {author}
  {\bibfnamefont {M.~G.}\ \bibnamefont {Lagally}}, \bibinfo {author}
  {\bibfnamefont {R.~H.}\ \bibnamefont {Foote}}, \bibinfo {author}
  {\bibfnamefont {M.}~\bibnamefont {Friesen}}, \bibinfo {author} {\bibfnamefont
  {S.~N.}\ \bibnamefont {Coppersmith}}, \ and\ \bibinfo {author} {\bibfnamefont
  {M.~A.}\ \bibnamefont {Eriksson}},\ }\href {\doibase 10.1038/npjqi.2016.32}
  {\bibfield  {journal} {\bibinfo  {journal} {npj Quantum Inf.}\ }\textbf
  {\bibinfo {volume} {2}},\ \bibinfo {pages} {16032} (\bibinfo {year}
  {2016})}\BibitemShut {NoStop}%
\bibitem [{\citenamefont {Yang}\ \emph {et~al.}(2019)\citenamefont {Yang},
  \citenamefont {Coppersmith},\ and\ \citenamefont {Friesen}}]{Yang.19b}%
  \BibitemOpen
  \bibfield  {author} {\bibinfo {author} {\bibfnamefont {Y.-C.}\ \bibnamefont
  {Yang}}, \bibinfo {author} {\bibfnamefont {S.}~\bibnamefont {Coppersmith}}, \
  and\ \bibinfo {author} {\bibfnamefont {M.}~\bibnamefont {Friesen}},\ }\href
  {\doibase 10.1103/PhysRevA.100.022337} {\bibfield  {journal} {\bibinfo
  {journal} {Phys. Rev. A}\ }\textbf {\bibinfo {volume} {100}},\ \bibinfo
  {pages} {022337} (\bibinfo {year} {2019})}\BibitemShut {NoStop}%
\bibitem [{\citenamefont {Noiri}\ \emph {et~al.}(2022)\citenamefont {Noiri},
  \citenamefont {Takeda}, \citenamefont {Nakajima}, \citenamefont {Kobayashi},
  \citenamefont {Sammak}, \citenamefont {Scappucci},\ and\ \citenamefont
  {Tarucha}}]{Noiri.22}%
  \BibitemOpen
  \bibfield  {author} {\bibinfo {author} {\bibfnamefont {A.}~\bibnamefont
  {Noiri}}, \bibinfo {author} {\bibfnamefont {K.}~\bibnamefont {Takeda}},
  \bibinfo {author} {\bibfnamefont {T.}~\bibnamefont {Nakajima}}, \bibinfo
  {author} {\bibfnamefont {T.}~\bibnamefont {Kobayashi}}, \bibinfo {author}
  {\bibfnamefont {A.}~\bibnamefont {Sammak}}, \bibinfo {author} {\bibfnamefont
  {G.}~\bibnamefont {Scappucci}}, \ and\ \bibinfo {author} {\bibfnamefont
  {S.}~\bibnamefont {Tarucha}},\ }\href {\doibase 10.1038/s41586-021-04182-y}
  {\bibfield  {journal} {\bibinfo  {journal} {Nature}\ }\textbf {\bibinfo
  {volume} {601}},\ \bibinfo {pages} {338} (\bibinfo {year}
  {2022})}\BibitemShut {NoStop}%
\bibitem [{\citenamefont {Xue}\ \emph {et~al.}(2022)\citenamefont {Xue},
  \citenamefont {Russ}, \citenamefont {Samkharadze}, \citenamefont {Undseth},
  \citenamefont {Sammak}, \citenamefont {Scappucci},\ and\ \citenamefont
  {Vandersypen}}]{Xue.22}%
  \BibitemOpen
  \bibfield  {author} {\bibinfo {author} {\bibfnamefont {X.}~\bibnamefont
  {Xue}}, \bibinfo {author} {\bibfnamefont {M.}~\bibnamefont {Russ}}, \bibinfo
  {author} {\bibfnamefont {N.}~\bibnamefont {Samkharadze}}, \bibinfo {author}
  {\bibfnamefont {B.}~\bibnamefont {Undseth}}, \bibinfo {author} {\bibfnamefont
  {A.}~\bibnamefont {Sammak}}, \bibinfo {author} {\bibfnamefont
  {G.}~\bibnamefont {Scappucci}}, \ and\ \bibinfo {author} {\bibfnamefont
  {L.~M.~K.}\ \bibnamefont {Vandersypen}},\ }\href {\doibase
  10.1038/s41586-021-04273-w} {\bibfield  {journal} {\bibinfo  {journal}
  {Nature}\ }\textbf {\bibinfo {volume} {601}},\ \bibinfo {pages} {343}
  (\bibinfo {year} {2022})}\BibitemShut {NoStop}%
\bibitem [{\citenamefont {Madzik}\ \emph {et~al.}(2022)\citenamefont {Madzik},
  \citenamefont {Asaad}, \citenamefont {Youssry}, \citenamefont {Joecker},
  \citenamefont {Rudinger}, \citenamefont {Nielsen}, \citenamefont {Young},
  \citenamefont {Proctor}, \citenamefont {Baczewski}, \citenamefont {Laucht},
  \citenamefont {Schmitt}, \citenamefont {Hudson}, \citenamefont {Itoh},
  \citenamefont {Jakob}, \citenamefont {Johnson}, \citenamefont {Jamieson},
  \citenamefont {Dzurak}, \citenamefont {Ferrie}, \citenamefont
  {Blume-Kohout},\ and\ \citenamefont {Morello}}]{Madzik.22}%
  \BibitemOpen
  \bibfield  {author} {\bibinfo {author} {\bibfnamefont {M.~T.}\ \bibnamefont
  {Madzik}}, \bibinfo {author} {\bibfnamefont {S.}~\bibnamefont {Asaad}},
  \bibinfo {author} {\bibfnamefont {A.}~\bibnamefont {Youssry}}, \bibinfo
  {author} {\bibfnamefont {B.}~\bibnamefont {Joecker}}, \bibinfo {author}
  {\bibfnamefont {K.~M.}\ \bibnamefont {Rudinger}}, \bibinfo {author}
  {\bibfnamefont {E.}~\bibnamefont {Nielsen}}, \bibinfo {author} {\bibfnamefont
  {K.~C.}\ \bibnamefont {Young}}, \bibinfo {author} {\bibfnamefont {T.~J.}\
  \bibnamefont {Proctor}}, \bibinfo {author} {\bibfnamefont {A.~D.}\
  \bibnamefont {Baczewski}}, \bibinfo {author} {\bibfnamefont {A.}~\bibnamefont
  {Laucht}}, \bibinfo {author} {\bibfnamefont {V.}~\bibnamefont {Schmitt}},
  \bibinfo {author} {\bibfnamefont {F.~E.}\ \bibnamefont {Hudson}}, \bibinfo
  {author} {\bibfnamefont {K.~M.}\ \bibnamefont {Itoh}}, \bibinfo {author}
  {\bibfnamefont {A.~M.}\ \bibnamefont {Jakob}}, \bibinfo {author}
  {\bibfnamefont {B.~C.}\ \bibnamefont {Johnson}}, \bibinfo {author}
  {\bibfnamefont {D.~N.}\ \bibnamefont {Jamieson}}, \bibinfo {author}
  {\bibfnamefont {A.~S.}\ \bibnamefont {Dzurak}}, \bibinfo {author}
  {\bibfnamefont {C.}~\bibnamefont {Ferrie}}, \bibinfo {author} {\bibfnamefont
  {R.}~\bibnamefont {Blume-Kohout}}, \ and\ \bibinfo {author} {\bibfnamefont
  {A.}~\bibnamefont {Morello}},\ }\href {\doibase 10.1038/s41586-021-04292-7}
  {\bibfield  {journal} {\bibinfo  {journal} {Nature}\ }\textbf {\bibinfo
  {volume} {601}},\ \bibinfo {pages} {348} (\bibinfo {year}
  {2022})}\BibitemShut {NoStop}%
\bibitem [{\citenamefont {Dial}\ \emph {et~al.}(2013)\citenamefont {Dial},
  \citenamefont {Shulman}, \citenamefont {Harvey}, \citenamefont {Bluhm},
  \citenamefont {Umansky},\ and\ \citenamefont {Yacoby}}]{Dial.13}%
  \BibitemOpen
  \bibfield  {author} {\bibinfo {author} {\bibfnamefont {O.~E.}\ \bibnamefont
  {Dial}}, \bibinfo {author} {\bibfnamefont {M.~D.}\ \bibnamefont {Shulman}},
  \bibinfo {author} {\bibfnamefont {S.~P.}\ \bibnamefont {Harvey}}, \bibinfo
  {author} {\bibfnamefont {H.}~\bibnamefont {Bluhm}}, \bibinfo {author}
  {\bibfnamefont {V.}~\bibnamefont {Umansky}}, \ and\ \bibinfo {author}
  {\bibfnamefont {A.}~\bibnamefont {Yacoby}},\ }\href {\doibase
  10.1103/PhysRevLett.110.146804} {\bibfield  {journal} {\bibinfo  {journal}
  {Phys. Rev. Lett.}\ }\textbf {\bibinfo {volume} {110}},\ \bibinfo {pages}
  {146804} (\bibinfo {year} {2013})}\BibitemShut {NoStop}%
\bibitem [{\citenamefont {van Woerkom}\ \emph {et~al.}(2018)\citenamefont {van
  Woerkom}, \citenamefont {Scarlino}, \citenamefont {Ungerer}, \citenamefont
  {M\"uller}, \citenamefont {Koski}, \citenamefont {Landig}, \citenamefont
  {Reichl}, \citenamefont {Wegscheider}, \citenamefont {Ihn}, \citenamefont
  {Ensslin},\ and\ \citenamefont {Wallraff}}]{Van.18}%
  \BibitemOpen
  \bibfield  {author} {\bibinfo {author} {\bibfnamefont {D.~J.}\ \bibnamefont
  {van Woerkom}}, \bibinfo {author} {\bibfnamefont {P.}~\bibnamefont
  {Scarlino}}, \bibinfo {author} {\bibfnamefont {J.~H.}\ \bibnamefont
  {Ungerer}}, \bibinfo {author} {\bibfnamefont {C.}~\bibnamefont {M\"uller}},
  \bibinfo {author} {\bibfnamefont {J.~V.}\ \bibnamefont {Koski}}, \bibinfo
  {author} {\bibfnamefont {A.~J.}\ \bibnamefont {Landig}}, \bibinfo {author}
  {\bibfnamefont {C.}~\bibnamefont {Reichl}}, \bibinfo {author} {\bibfnamefont
  {W.}~\bibnamefont {Wegscheider}}, \bibinfo {author} {\bibfnamefont
  {T.}~\bibnamefont {Ihn}}, \bibinfo {author} {\bibfnamefont {K.}~\bibnamefont
  {Ensslin}}, \ and\ \bibinfo {author} {\bibfnamefont {A.}~\bibnamefont
  {Wallraff}},\ }\href {\doibase 10.1103/PhysRevX.8.041018} {\bibfield
  {journal} {\bibinfo  {journal} {Phys. Rev. X}\ }\textbf {\bibinfo {volume}
  {8}},\ \bibinfo {pages} {041018} (\bibinfo {year} {2018})}\BibitemShut
  {NoStop}%
\bibitem [{\citenamefont {MacQuarrie}\ \emph {et~al.}(2020)\citenamefont
  {MacQuarrie}, \citenamefont {Neyens}, \citenamefont {Dodson}, \citenamefont
  {Corrigan}, \citenamefont {Thorgrimsson}, \citenamefont {Holman},
  \citenamefont {Palma}, \citenamefont {Edge}, \citenamefont {Friesen},
  \citenamefont {Coppersmith},\ and\ \citenamefont {Eriksson}}]{macquarrie.20}%
  \BibitemOpen
  \bibfield  {author} {\bibinfo {author} {\bibfnamefont {E.~R.}\ \bibnamefont
  {MacQuarrie}}, \bibinfo {author} {\bibfnamefont {S.~F.}\ \bibnamefont
  {Neyens}}, \bibinfo {author} {\bibfnamefont {J.~P.}\ \bibnamefont {Dodson}},
  \bibinfo {author} {\bibfnamefont {J.}~\bibnamefont {Corrigan}}, \bibinfo
  {author} {\bibfnamefont {B.}~\bibnamefont {Thorgrimsson}}, \bibinfo {author}
  {\bibfnamefont {N.}~\bibnamefont {Holman}}, \bibinfo {author} {\bibfnamefont
  {M.}~\bibnamefont {Palma}}, \bibinfo {author} {\bibfnamefont {L.~F.}\
  \bibnamefont {Edge}}, \bibinfo {author} {\bibfnamefont {M.}~\bibnamefont
  {Friesen}}, \bibinfo {author} {\bibfnamefont {S.~N.}\ \bibnamefont
  {Coppersmith}}, \ and\ \bibinfo {author} {\bibfnamefont {M.~A.}\ \bibnamefont
  {Eriksson}},\ }\href {https://www.nature.com/articles/s41534-020-00314-w}
  {\bibfield  {journal} {\bibinfo  {journal} {npj Quantum Inf.}\ }\textbf
  {\bibinfo {volume} {6}},\ \bibinfo {pages} {81} (\bibinfo {year}
  {2020})}\BibitemShut {NoStop}%
\bibitem [{\citenamefont {Friesen}\ \emph {et~al.}(2017)\citenamefont
  {Friesen}, \citenamefont {Ghosh}, \citenamefont {Eriksson},\ and\
  \citenamefont {Coppersmith}}]{Friesen.17}%
  \BibitemOpen
  \bibfield  {author} {\bibinfo {author} {\bibfnamefont {M.}~\bibnamefont
  {Friesen}}, \bibinfo {author} {\bibfnamefont {J.}~\bibnamefont {Ghosh}},
  \bibinfo {author} {\bibfnamefont {M.}~\bibnamefont {Eriksson}}, \ and\
  \bibinfo {author} {\bibfnamefont {S.}~\bibnamefont {Coppersmith}},\ }\href
  {\doibase 10.1038/ncomms15923 (2017)} {\bibfield  {journal} {\bibinfo
  {journal} {Nat. Commun}\ }\textbf {\bibinfo {volume} {8}},\ \bibinfo {pages}
  {15923} (\bibinfo {year} {2017})}\BibitemShut {NoStop}%
\bibitem [{\citenamefont {Koski}\ \emph {et~al.}(2020)\citenamefont {Koski},
  \citenamefont {Landig}, \citenamefont {Russ}, \citenamefont {Abadillo-Uriel},
  \citenamefont {Scarlino}, \citenamefont {Kratochwil}, \citenamefont {Reichl},
  \citenamefont {Wegscheider}, \citenamefont {Burkard}, \citenamefont
  {Friesen}, \citenamefont {Coppersmith}, \citenamefont {Wallraff},
  \citenamefont {Ensslin},\ and\ \citenamefont {Ihn}}]{koski.20}%
  \BibitemOpen
  \bibfield  {author} {\bibinfo {author} {\bibfnamefont {J.}~\bibnamefont
  {Koski}, \bibfnamefont {V}}, \bibinfo {author} {\bibfnamefont {A.~J.}\
  \bibnamefont {Landig}}, \bibinfo {author} {\bibfnamefont {M.}~\bibnamefont
  {Russ}}, \bibinfo {author} {\bibfnamefont {J.~C.}\ \bibnamefont
  {Abadillo-Uriel}}, \bibinfo {author} {\bibfnamefont {P.}~\bibnamefont
  {Scarlino}}, \bibinfo {author} {\bibfnamefont {B.}~\bibnamefont
  {Kratochwil}}, \bibinfo {author} {\bibfnamefont {C.}~\bibnamefont {Reichl}},
  \bibinfo {author} {\bibfnamefont {W.}~\bibnamefont {Wegscheider}}, \bibinfo
  {author} {\bibfnamefont {G.}~\bibnamefont {Burkard}}, \bibinfo {author}
  {\bibfnamefont {M.}~\bibnamefont {Friesen}}, \bibinfo {author} {\bibfnamefont
  {S.~N.}\ \bibnamefont {Coppersmith}}, \bibinfo {author} {\bibfnamefont
  {A.}~\bibnamefont {Wallraff}}, \bibinfo {author} {\bibfnamefont
  {K.}~\bibnamefont {Ensslin}}, \ and\ \bibinfo {author} {\bibfnamefont
  {T.}~\bibnamefont {Ihn}},\ }\href {\doibase 10.1038/s41567-020-0862-4}
  {\bibfield  {journal} {\bibinfo  {journal} {Nat. Phys.}\ }\textbf {\bibinfo
  {volume} {16}},\ \bibinfo {pages} {642} (\bibinfo {year} {2020})}\BibitemShut
  {NoStop}%
\bibitem [{\citenamefont {Ghosh}\ \emph {et~al.}(2017)\citenamefont {Ghosh},
  \citenamefont {Coppersmith},\ and\ \citenamefont {Friesen}}]{Ghosh.17}%
  \BibitemOpen
  \bibfield  {author} {\bibinfo {author} {\bibfnamefont {J.}~\bibnamefont
  {Ghosh}}, \bibinfo {author} {\bibfnamefont {S.~N.}\ \bibnamefont
  {Coppersmith}}, \ and\ \bibinfo {author} {\bibfnamefont {M.}~\bibnamefont
  {Friesen}},\ }\href {\doibase 10.1103/PhysRevB.95.241307} {\bibfield
  {journal} {\bibinfo  {journal} {Phys. Rev. B}\ }\textbf {\bibinfo {volume}
  {95}},\ \bibinfo {pages} {241307} (\bibinfo {year} {2017})}\BibitemShut
  {NoStop}%
\bibitem [{\citenamefont {Blais}\ \emph {et~al.}(2007)\citenamefont {Blais},
  \citenamefont {Gambetta}, \citenamefont {Wallraff}, \citenamefont {Schuster},
  \citenamefont {Girvin}, \citenamefont {Devoret},\ and\ \citenamefont
  {Schoelkopf}}]{blais.07}%
  \BibitemOpen
  \bibfield  {author} {\bibinfo {author} {\bibfnamefont {A.}~\bibnamefont
  {Blais}}, \bibinfo {author} {\bibfnamefont {J.}~\bibnamefont {Gambetta}},
  \bibinfo {author} {\bibfnamefont {A.}~\bibnamefont {Wallraff}}, \bibinfo
  {author} {\bibfnamefont {D.~I.}\ \bibnamefont {Schuster}}, \bibinfo {author}
  {\bibfnamefont {S.~M.}\ \bibnamefont {Girvin}}, \bibinfo {author}
  {\bibfnamefont {M.~H.}\ \bibnamefont {Devoret}}, \ and\ \bibinfo {author}
  {\bibfnamefont {R.~J.}\ \bibnamefont {Schoelkopf}},\ }\href
  {https://link.aps.org/doi/10.1103/PhysRevA.75.032329} {\bibfield  {journal}
  {\bibinfo  {journal} {Phys. Rev. A}\ }\textbf {\bibinfo {volume} {75}},\
  \bibinfo {pages} {032329} (\bibinfo {year} {2007})}\BibitemShut {NoStop}%
\bibitem [{\citenamefont {Srinivasa}\ \emph {et~al.}(2016)\citenamefont
  {Srinivasa}, \citenamefont {Taylor},\ and\ \citenamefont
  {Tahan}}]{Srinivasa.16}%
  \BibitemOpen
  \bibfield  {author} {\bibinfo {author} {\bibfnamefont {V.}~\bibnamefont
  {Srinivasa}}, \bibinfo {author} {\bibfnamefont {J.~M.}\ \bibnamefont
  {Taylor}}, \ and\ \bibinfo {author} {\bibfnamefont {C.}~\bibnamefont
  {Tahan}},\ }\href {\doibase 10.1103/PhysRevB.94.205421} {\bibfield  {journal}
  {\bibinfo  {journal} {Phys. Rev. B}\ }\textbf {\bibinfo {volume} {94}},\
  \bibinfo {pages} {205421} (\bibinfo {year} {2016})}\BibitemShut {NoStop}%
\bibitem [{\citenamefont {Scarlino}\ \emph {et~al.}(2019)\citenamefont
  {Scarlino}, \citenamefont {van Woerkom}, \citenamefont {Mendes},
  \citenamefont {Koski}, \citenamefont {Landig}, \citenamefont {Andersen},
  \citenamefont {Gasparinetti}, \citenamefont {Reichl}, \citenamefont
  {Wegscheider}, \citenamefont {Ensslin}, \citenamefont {Ihn}, \citenamefont
  {Blais},\ and\ \citenamefont {Wallraff}}]{scarlino.19}%
  \BibitemOpen
  \bibfield  {author} {\bibinfo {author} {\bibfnamefont {P.}~\bibnamefont
  {Scarlino}}, \bibinfo {author} {\bibfnamefont {D.~J.}\ \bibnamefont {van
  Woerkom}}, \bibinfo {author} {\bibfnamefont {U.~C.}\ \bibnamefont {Mendes}},
  \bibinfo {author} {\bibfnamefont {J.~V.}\ \bibnamefont {Koski}}, \bibinfo
  {author} {\bibfnamefont {A.~J.}\ \bibnamefont {Landig}}, \bibinfo {author}
  {\bibfnamefont {C.~K.}\ \bibnamefont {Andersen}}, \bibinfo {author}
  {\bibfnamefont {S.}~\bibnamefont {Gasparinetti}}, \bibinfo {author}
  {\bibfnamefont {C.}~\bibnamefont {Reichl}}, \bibinfo {author} {\bibfnamefont
  {W.}~\bibnamefont {Wegscheider}}, \bibinfo {author} {\bibfnamefont
  {K.}~\bibnamefont {Ensslin}}, \bibinfo {author} {\bibfnamefont
  {T.}~\bibnamefont {Ihn}}, \bibinfo {author} {\bibfnamefont {A.}~\bibnamefont
  {Blais}}, \ and\ \bibinfo {author} {\bibfnamefont {A.}~\bibnamefont
  {Wallraff}},\ }\href {\doibase 10.1038/s41467-019-10798-6} {\bibfield
  {journal} {\bibinfo  {journal} {Nat. Commun.}\ }\textbf {\bibinfo {volume}
  {10}},\ \bibinfo {pages} {3011} (\bibinfo {year} {2019})}\BibitemShut
  {NoStop}%
\bibitem [{\citenamefont {Landig}\ \emph {et~al.}(2019)\citenamefont {Landig},
  \citenamefont {Koski}, \citenamefont {Scarlino}, \citenamefont {M{\"u}ller},
  \citenamefont {Abadillo-Uriel}, \citenamefont {Kratochwil}, \citenamefont
  {Reichl}, \citenamefont {Wegscheider}, \citenamefont {Coppersmith},
  \citenamefont {Friesen} \emph {et~al.}}]{landig.19}%
  \BibitemOpen
  \bibfield  {author} {\bibinfo {author} {\bibfnamefont {A.~J.}\ \bibnamefont
  {Landig}}, \bibinfo {author} {\bibfnamefont {J.~V.}\ \bibnamefont {Koski}},
  \bibinfo {author} {\bibfnamefont {P.}~\bibnamefont {Scarlino}}, \bibinfo
  {author} {\bibfnamefont {C.}~\bibnamefont {M{\"u}ller}}, \bibinfo {author}
  {\bibfnamefont {J.~C.}\ \bibnamefont {Abadillo-Uriel}}, \bibinfo {author}
  {\bibfnamefont {B.}~\bibnamefont {Kratochwil}}, \bibinfo {author}
  {\bibfnamefont {C.}~\bibnamefont {Reichl}}, \bibinfo {author} {\bibfnamefont
  {W.}~\bibnamefont {Wegscheider}}, \bibinfo {author} {\bibfnamefont {S.~N.}\
  \bibnamefont {Coppersmith}}, \bibinfo {author} {\bibfnamefont
  {M.}~\bibnamefont {Friesen}},  \emph {et~al.},\ }\href {\doibase
  10.1038/s41467-019-13000-z} {\bibfield  {journal} {\bibinfo  {journal} {Nat.
  Commun.}\ }\textbf {\bibinfo {volume} {10}},\ \bibinfo {pages} {5037}
  (\bibinfo {year} {2019})}\BibitemShut {NoStop}%
\bibitem [{\citenamefont {Wang}\ \emph {et~al.}(2014)\citenamefont {Wang},
  \citenamefont {Bishop}, \citenamefont {Barnes}, \citenamefont {Kestner},\
  and\ \citenamefont {Sarma}}]{Wang.14}%
  \BibitemOpen
  \bibfield  {author} {\bibinfo {author} {\bibfnamefont {X.}~\bibnamefont
  {Wang}}, \bibinfo {author} {\bibfnamefont {L.~S.}\ \bibnamefont {Bishop}},
  \bibinfo {author} {\bibfnamefont {E.}~\bibnamefont {Barnes}}, \bibinfo
  {author} {\bibfnamefont {J.~P.}\ \bibnamefont {Kestner}}, \ and\ \bibinfo
  {author} {\bibfnamefont {S.~D.}\ \bibnamefont {Sarma}},\ }\href {\doibase
  10.1103/PhysRevA.89.022310} {\bibfield  {journal} {\bibinfo  {journal} {Phys.
  Rev. A}\ }\textbf {\bibinfo {volume} {89}},\ \bibinfo {pages} {022310}
  (\bibinfo {year} {2014})}\BibitemShut {NoStop}%
\bibitem [{\citenamefont {Kratochwil}\ \emph {et~al.}(2021)\citenamefont
  {Kratochwil}, \citenamefont {Koski}, \citenamefont {Landig}, \citenamefont
  {Scarlino}, \citenamefont {Abadillo-Uriel}, \citenamefont {Reichl},
  \citenamefont {Coppersmith}, \citenamefont {Wegscheider}, \citenamefont
  {Friesen}, \citenamefont {Wallraff}, \citenamefont {Ihn},\ and\ \citenamefont
  {Ensslin}}]{Kratochwil.21}%
  \BibitemOpen
  \bibfield  {author} {\bibinfo {author} {\bibfnamefont {B.}~\bibnamefont
  {Kratochwil}}, \bibinfo {author} {\bibfnamefont {J.~V.}\ \bibnamefont
  {Koski}}, \bibinfo {author} {\bibfnamefont {A.~J.}\ \bibnamefont {Landig}},
  \bibinfo {author} {\bibfnamefont {P.}~\bibnamefont {Scarlino}}, \bibinfo
  {author} {\bibfnamefont {J.~C.}\ \bibnamefont {Abadillo-Uriel}}, \bibinfo
  {author} {\bibfnamefont {C.}~\bibnamefont {Reichl}}, \bibinfo {author}
  {\bibfnamefont {S.~N.}\ \bibnamefont {Coppersmith}}, \bibinfo {author}
  {\bibfnamefont {W.}~\bibnamefont {Wegscheider}}, \bibinfo {author}
  {\bibfnamefont {M.}~\bibnamefont {Friesen}}, \bibinfo {author} {\bibfnamefont
  {A.}~\bibnamefont {Wallraff}}, \bibinfo {author} {\bibfnamefont
  {T.}~\bibnamefont {Ihn}}, \ and\ \bibinfo {author} {\bibfnamefont
  {K.}~\bibnamefont {Ensslin}},\ }\href {\doibase
  10.1103/PhysRevResearch.3.013171} {\bibfield  {journal} {\bibinfo  {journal}
  {Phys. Rev. Res.}\ }\textbf {\bibinfo {volume} {3}},\ \bibinfo {pages}
  {013171} (\bibinfo {year} {2021})}\BibitemShut {NoStop}%
\bibitem [{\citenamefont {Russ}\ \emph {et~al.}(2018)\citenamefont {Russ},
  \citenamefont {Zajac}, \citenamefont {Sigillito}, \citenamefont {Borjans},
  \citenamefont {Taylor}, \citenamefont {Petta},\ and\ \citenamefont
  {Burkard}}]{Russ.18}%
  \BibitemOpen
  \bibfield  {author} {\bibinfo {author} {\bibfnamefont {M.}~\bibnamefont
  {Russ}}, \bibinfo {author} {\bibfnamefont {D.~M.}\ \bibnamefont {Zajac}},
  \bibinfo {author} {\bibfnamefont {A.~J.}\ \bibnamefont {Sigillito}}, \bibinfo
  {author} {\bibfnamefont {F.}~\bibnamefont {Borjans}}, \bibinfo {author}
  {\bibfnamefont {J.~M.}\ \bibnamefont {Taylor}}, \bibinfo {author}
  {\bibfnamefont {J.~R.}\ \bibnamefont {Petta}}, \ and\ \bibinfo {author}
  {\bibfnamefont {G.}~\bibnamefont {Burkard}},\ }\href {\doibase
  10.1103/PhysRevB.97.085421} {\bibfield  {journal} {\bibinfo  {journal} {Phys.
  Rev. B}\ }\textbf {\bibinfo {volume} {97}},\ \bibinfo {pages} {085421}
  (\bibinfo {year} {2018})}\BibitemShut {NoStop}%
\bibitem [{\citenamefont {Childress}\ \emph {et~al.}(2004)\citenamefont
  {Childress}, \citenamefont {S\o{}rensen},\ and\ \citenamefont
  {Lukin}}]{Childress.04}%
  \BibitemOpen
  \bibfield  {author} {\bibinfo {author} {\bibfnamefont {L.}~\bibnamefont
  {Childress}}, \bibinfo {author} {\bibfnamefont {A.~S.}\ \bibnamefont
  {S\o{}rensen}}, \ and\ \bibinfo {author} {\bibfnamefont {M.~D.}\ \bibnamefont
  {Lukin}},\ }\href {\doibase 10.1103/PhysRevA.69.042302} {\bibfield  {journal}
  {\bibinfo  {journal} {Phys. Rev. A}\ }\textbf {\bibinfo {volume} {69}},\
  \bibinfo {pages} {042302} (\bibinfo {year} {2004})}\BibitemShut {NoStop}%
\bibitem [{\citenamefont {Blais}\ \emph {et~al.}(2004)\citenamefont {Blais},
  \citenamefont {Huang}, \citenamefont {Wallraff}, \citenamefont {Girvin},\
  and\ \citenamefont {Schoelkopf}}]{Blais.04}%
  \BibitemOpen
  \bibfield  {author} {\bibinfo {author} {\bibfnamefont {A.}~\bibnamefont
  {Blais}}, \bibinfo {author} {\bibfnamefont {R.-S.}\ \bibnamefont {Huang}},
  \bibinfo {author} {\bibfnamefont {A.}~\bibnamefont {Wallraff}}, \bibinfo
  {author} {\bibfnamefont {S.~M.}\ \bibnamefont {Girvin}}, \ and\ \bibinfo
  {author} {\bibfnamefont {R.~J.}\ \bibnamefont {Schoelkopf}},\ }\href
  {\doibase 10.1103/PhysRevA.69.062320} {\bibfield  {journal} {\bibinfo
  {journal} {Phys. Rev. A}\ }\textbf {\bibinfo {volume} {69}},\ \bibinfo
  {pages} {062320} (\bibinfo {year} {2004})}\BibitemShut {NoStop}%
\bibitem [{\citenamefont {Stockklauser}\ \emph {et~al.}(2017)\citenamefont
  {Stockklauser}, \citenamefont {Scarlino}, \citenamefont {Koski},
  \citenamefont {Gasparinetti}, \citenamefont {Andersen}, \citenamefont
  {Reichl}, \citenamefont {Wegscheider}, \citenamefont {Ihn}, \citenamefont
  {Ensslin},\ and\ \citenamefont {Wallraff}}]{Stockklauser.17}%
  \BibitemOpen
  \bibfield  {author} {\bibinfo {author} {\bibfnamefont {A.}~\bibnamefont
  {Stockklauser}}, \bibinfo {author} {\bibfnamefont {P.}~\bibnamefont
  {Scarlino}}, \bibinfo {author} {\bibfnamefont {J.~V.}\ \bibnamefont {Koski}},
  \bibinfo {author} {\bibfnamefont {S.}~\bibnamefont {Gasparinetti}}, \bibinfo
  {author} {\bibfnamefont {C.~K.}\ \bibnamefont {Andersen}}, \bibinfo {author}
  {\bibfnamefont {C.}~\bibnamefont {Reichl}}, \bibinfo {author} {\bibfnamefont
  {W.}~\bibnamefont {Wegscheider}}, \bibinfo {author} {\bibfnamefont
  {T.}~\bibnamefont {Ihn}}, \bibinfo {author} {\bibfnamefont {K.}~\bibnamefont
  {Ensslin}}, \ and\ \bibinfo {author} {\bibfnamefont {A.}~\bibnamefont
  {Wallraff}},\ }\href {\doibase 10.1103/PhysRevX.7.011030} {\bibfield
  {journal} {\bibinfo  {journal} {Phys. Rev. X}\ }\textbf {\bibinfo {volume}
  {7}},\ \bibinfo {pages} {011030} (\bibinfo {year} {2017})}\BibitemShut
  {NoStop}%
\bibitem [{\citenamefont {Wang}\ \emph {et~al.}(2021)\citenamefont {Wang},
  \citenamefont {Lin}, \citenamefont {Li}, \citenamefont {Gu}, \citenamefont
  {Chen}, \citenamefont {Guo}, \citenamefont {Jiang}, \citenamefont {Hu},
  \citenamefont {Cao},\ and\ \citenamefont {Guo}}]{Wang.20}%
  \BibitemOpen
  \bibfield  {author} {\bibinfo {author} {\bibfnamefont {B.-C.}\ \bibnamefont
  {Wang}}, \bibinfo {author} {\bibfnamefont {T.}~\bibnamefont {Lin}}, \bibinfo
  {author} {\bibfnamefont {H.-O.}\ \bibnamefont {Li}}, \bibinfo {author}
  {\bibfnamefont {S.-S.}\ \bibnamefont {Gu}}, \bibinfo {author} {\bibfnamefont
  {M.-B.}\ \bibnamefont {Chen}}, \bibinfo {author} {\bibfnamefont {G.-C.}\
  \bibnamefont {Guo}}, \bibinfo {author} {\bibfnamefont {H.-W.}\ \bibnamefont
  {Jiang}}, \bibinfo {author} {\bibfnamefont {X.}~\bibnamefont {Hu}}, \bibinfo
  {author} {\bibfnamefont {G.}~\bibnamefont {Cao}}, \ and\ \bibinfo {author}
  {\bibfnamefont {G.-P.}\ \bibnamefont {Guo}},\ }\href {\doibase
  https://doi.org/10.1016/j.scib.2020.10.005} {\bibfield  {journal} {\bibinfo
  {journal} {Sci. Bull.}\ }\textbf {\bibinfo {volume} {66}},\ \bibinfo {pages}
  {332} (\bibinfo {year} {2021})}\BibitemShut {NoStop}%
\bibitem [{\citenamefont {Fink}\ \emph {et~al.}(2009)\citenamefont {Fink},
  \citenamefont {Bianchetti}, \citenamefont {Baur}, \citenamefont {G\"oppl},
  \citenamefont {Steffen}, \citenamefont {Filipp}, \citenamefont {Leek},
  \citenamefont {Blais},\ and\ \citenamefont {Wallraff}}]{Fink.09}%
  \BibitemOpen
  \bibfield  {author} {\bibinfo {author} {\bibfnamefont {J.~M.}\ \bibnamefont
  {Fink}}, \bibinfo {author} {\bibfnamefont {R.}~\bibnamefont {Bianchetti}},
  \bibinfo {author} {\bibfnamefont {M.}~\bibnamefont {Baur}}, \bibinfo {author}
  {\bibfnamefont {M.}~\bibnamefont {G\"oppl}}, \bibinfo {author} {\bibfnamefont
  {L.}~\bibnamefont {Steffen}}, \bibinfo {author} {\bibfnamefont
  {S.}~\bibnamefont {Filipp}}, \bibinfo {author} {\bibfnamefont {P.~J.}\
  \bibnamefont {Leek}}, \bibinfo {author} {\bibfnamefont {A.}~\bibnamefont
  {Blais}}, \ and\ \bibinfo {author} {\bibfnamefont {A.}~\bibnamefont
  {Wallraff}},\ }\href {\doibase 10.1103/PhysRevLett.103.083601} {\bibfield
  {journal} {\bibinfo  {journal} {Phys. Rev. Lett.}\ }\textbf {\bibinfo
  {volume} {103}},\ \bibinfo {pages} {083601} (\bibinfo {year}
  {2009})}\BibitemShut {NoStop}%
\bibitem [{\citenamefont {Hong}\ \emph {et~al.}(2018)\citenamefont {Hong},
  \citenamefont {Liu}, \citenamefont {Cai}, \citenamefont {Zhang},
  \citenamefont {Hu}, \citenamefont {Wang},\ and\ \citenamefont
  {Xue}}]{hong.18}%
  \BibitemOpen
  \bibfield  {author} {\bibinfo {author} {\bibfnamefont {Z.-P.}\ \bibnamefont
  {Hong}}, \bibinfo {author} {\bibfnamefont {B.-J.}\ \bibnamefont {Liu}},
  \bibinfo {author} {\bibfnamefont {J.-Q.}\ \bibnamefont {Cai}}, \bibinfo
  {author} {\bibfnamefont {X.-D.}\ \bibnamefont {Zhang}}, \bibinfo {author}
  {\bibfnamefont {Y.}~\bibnamefont {Hu}}, \bibinfo {author} {\bibfnamefont
  {Z.~D.}\ \bibnamefont {Wang}}, \ and\ \bibinfo {author} {\bibfnamefont
  {Z.-Y.}\ \bibnamefont {Xue}},\ }\href {\doibase 10.1103/PhysRevA.97.022332}
  {\bibfield  {journal} {\bibinfo  {journal} {Phys. Rev. A}\ }\textbf {\bibinfo
  {volume} {97}},\ \bibinfo {pages} {022332} (\bibinfo {year}
  {2018})}\BibitemShut {NoStop}%
\bibitem [{\citenamefont {Egger}\ \emph {et~al.}(2019)\citenamefont {Egger},
  \citenamefont {Ganzhorn}, \citenamefont {Salis}, \citenamefont {Fuhrer},
  \citenamefont {M\"uller}, \citenamefont {Barkoutsos}, \citenamefont {Moll},
  \citenamefont {Tavernelli},\ and\ \citenamefont {Filipp}}]{Egger.19}%
  \BibitemOpen
  \bibfield  {author} {\bibinfo {author} {\bibfnamefont {D.}~\bibnamefont
  {Egger}}, \bibinfo {author} {\bibfnamefont {M.}~\bibnamefont {Ganzhorn}},
  \bibinfo {author} {\bibfnamefont {G.}~\bibnamefont {Salis}}, \bibinfo
  {author} {\bibfnamefont {A.}~\bibnamefont {Fuhrer}}, \bibinfo {author}
  {\bibfnamefont {P.}~\bibnamefont {M\"uller}}, \bibinfo {author}
  {\bibfnamefont {P.}~\bibnamefont {Barkoutsos}}, \bibinfo {author}
  {\bibfnamefont {N.}~\bibnamefont {Moll}}, \bibinfo {author} {\bibfnamefont
  {I.}~\bibnamefont {Tavernelli}}, \ and\ \bibinfo {author} {\bibfnamefont
  {S.}~\bibnamefont {Filipp}},\ }\href {\doibase
  10.1103/PhysRevApplied.11.014017} {\bibfield  {journal} {\bibinfo  {journal}
  {Phys. Rev. Appl.}\ }\textbf {\bibinfo {volume} {11}},\ \bibinfo {pages}
  {014017} (\bibinfo {year} {2019})}\BibitemShut {NoStop}%
\bibitem [{\citenamefont {Li}\ \emph {et~al.}(2020)\citenamefont {Li},
  \citenamefont {Chen},\ and\ \citenamefont {Xue}}]{li.20}%
  \BibitemOpen
  \bibfield  {author} {\bibinfo {author} {\bibfnamefont {S.}~\bibnamefont
  {Li}}, \bibinfo {author} {\bibfnamefont {T.}~\bibnamefont {Chen}}, \ and\
  \bibinfo {author} {\bibfnamefont {Z.-Y.}\ \bibnamefont {Xue}},\ }\href
  {\doibase https://doi.org/10.1002/qute.202000001} {\bibfield  {journal}
  {\bibinfo  {journal} {Adv. Quantum Technol.}\ }\textbf {\bibinfo {volume}
  {3}},\ \bibinfo {pages} {2000001} (\bibinfo {year} {2020})}\BibitemShut
  {NoStop}%
\bibitem [{\citenamefont {Zhang}\ \emph {et~al.}(2021)\citenamefont {Zhang},
  \citenamefont {Chen}, \citenamefont {Wang},\ and\ \citenamefont
  {Xue}}]{Zhang.20}%
  \BibitemOpen
  \bibfield  {author} {\bibinfo {author} {\bibfnamefont {C.}~\bibnamefont
  {Zhang}}, \bibinfo {author} {\bibfnamefont {T.}~\bibnamefont {Chen}},
  \bibinfo {author} {\bibfnamefont {X.}~\bibnamefont {Wang}}, \ and\ \bibinfo
  {author} {\bibfnamefont {Z.-Y.}\ \bibnamefont {Xue}},\ }\href
  {https://doi.org/10.1002/qute.202100011} {\bibfield  {journal} {\bibinfo
  {journal} {Adv. Quantum Technol.}\ }\textbf {\bibinfo {volume} {4}},\
  \bibinfo {pages} {2100011} (\bibinfo {year} {2021})}\BibitemShut {NoStop}%
\bibitem [{\citenamefont {Sj\"oqvist}\ \emph {et~al.}(2012)\citenamefont
  {Sj\"oqvist}, \citenamefont {Tong}, \citenamefont {Andersson}, \citenamefont
  {Hessmo}, \citenamefont {Johansson},\ and\ \citenamefont {Singh}}]{Erik.12}%
  \BibitemOpen
  \bibfield  {author} {\bibinfo {author} {\bibfnamefont {E.}~\bibnamefont
  {Sj\"oqvist}}, \bibinfo {author} {\bibfnamefont {D.~M.}\ \bibnamefont
  {Tong}}, \bibinfo {author} {\bibfnamefont {L.~M.}\ \bibnamefont {Andersson}},
  \bibinfo {author} {\bibfnamefont {B.}~\bibnamefont {Hessmo}}, \bibinfo
  {author} {\bibfnamefont {M.}~\bibnamefont {Johansson}}, \ and\ \bibinfo
  {author} {\bibfnamefont {K.}~\bibnamefont {Singh}},\ }\href
  {https://doi.org/10.1088%2F1367-2630%2F14%2F10%2F103035} {\bibfield
  {journal} {\bibinfo  {journal} {New J. Phys.}\ }\textbf {\bibinfo {volume}
  {14}},\ \bibinfo {pages} {103035} (\bibinfo {year} {2012})}\BibitemShut
  {NoStop}%
\bibitem [{\citenamefont {Sj{\"o}qvist}(2015)}]{Sjoqvist.15}%
  \BibitemOpen
  \bibfield  {author} {\bibinfo {author} {\bibfnamefont {E.}~\bibnamefont
  {Sj{\"o}qvist}},\ }\href {https://doi.org/10.1002/qua.24941} {\bibfield
  {journal} {\bibinfo  {journal} {Int. J. Quantum Chem.}\ }\textbf {\bibinfo
  {volume} {115}},\ \bibinfo {pages} {1311} (\bibinfo {year}
  {2015})}\BibitemShut {NoStop}%
\bibitem [{\citenamefont {Zhou}\ \emph {et~al.}(2018)\citenamefont {Zhou},
  \citenamefont {Liu}, \citenamefont {Hong},\ and\ \citenamefont
  {Xue}}]{Zhou.18}%
  \BibitemOpen
  \bibfield  {author} {\bibinfo {author} {\bibfnamefont {J.}~\bibnamefont
  {Zhou}}, \bibinfo {author} {\bibfnamefont {B.}~\bibnamefont {Liu}}, \bibinfo
  {author} {\bibfnamefont {Z.}~\bibnamefont {Hong}}, \ and\ \bibinfo {author}
  {\bibfnamefont {Z.}~\bibnamefont {Xue}},\ }\href {\doibase
  https://doi.org/10.1007/s11433-017-9119-8} {\bibfield  {journal} {\bibinfo
  {journal} {Sci. China Phys. Mech. Astron.}\ }\textbf {\bibinfo {volume}
  {61}},\ \bibinfo {pages} {010312} (\bibinfo {year} {2018})}\BibitemShut
  {NoStop}%
\bibitem [{\citenamefont {Chen}\ \emph {et~al.}(2016)\citenamefont {Chen},
  \citenamefont {Kelly}, \citenamefont {Quintana}, \citenamefont {Barends},
  \citenamefont {Campbell}, \citenamefont {Chen}, \citenamefont {Chiaro},
  \citenamefont {Dunsworth}, \citenamefont {Fowler}, \citenamefont {Lucero},
  \citenamefont {Jeffrey}, \citenamefont {Megrant}, \citenamefont {Mutus},
  \citenamefont {Neeley}, \citenamefont {Neill}, \citenamefont {O'Malley},
  \citenamefont {Roushan}, \citenamefont {Sank}, \citenamefont {Vainsencher},
  \citenamefont {Wenner}, \citenamefont {White}, \citenamefont {Korotkov},\
  and\ \citenamefont {Martinis}}]{Zi.16}%
  \BibitemOpen
  \bibfield  {author} {\bibinfo {author} {\bibfnamefont {Z.}~\bibnamefont
  {Chen}}, \bibinfo {author} {\bibfnamefont {J.}~\bibnamefont {Kelly}},
  \bibinfo {author} {\bibfnamefont {C.}~\bibnamefont {Quintana}}, \bibinfo
  {author} {\bibfnamefont {R.}~\bibnamefont {Barends}}, \bibinfo {author}
  {\bibfnamefont {B.}~\bibnamefont {Campbell}}, \bibinfo {author}
  {\bibfnamefont {Y.}~\bibnamefont {Chen}}, \bibinfo {author} {\bibfnamefont
  {B.}~\bibnamefont {Chiaro}}, \bibinfo {author} {\bibfnamefont
  {A.}~\bibnamefont {Dunsworth}}, \bibinfo {author} {\bibfnamefont {A.~G.}\
  \bibnamefont {Fowler}}, \bibinfo {author} {\bibfnamefont {E.}~\bibnamefont
  {Lucero}}, \bibinfo {author} {\bibfnamefont {E.}~\bibnamefont {Jeffrey}},
  \bibinfo {author} {\bibfnamefont {A.}~\bibnamefont {Megrant}}, \bibinfo
  {author} {\bibfnamefont {J.}~\bibnamefont {Mutus}}, \bibinfo {author}
  {\bibfnamefont {M.}~\bibnamefont {Neeley}}, \bibinfo {author} {\bibfnamefont
  {C.}~\bibnamefont {Neill}}, \bibinfo {author} {\bibfnamefont {P.~J.~J.}\
  \bibnamefont {O'Malley}}, \bibinfo {author} {\bibfnamefont {P.}~\bibnamefont
  {Roushan}}, \bibinfo {author} {\bibfnamefont {D.}~\bibnamefont {Sank}},
  \bibinfo {author} {\bibfnamefont {A.}~\bibnamefont {Vainsencher}}, \bibinfo
  {author} {\bibfnamefont {J.}~\bibnamefont {Wenner}}, \bibinfo {author}
  {\bibfnamefont {T.~C.}\ \bibnamefont {White}}, \bibinfo {author}
  {\bibfnamefont {A.~N.}\ \bibnamefont {Korotkov}}, \ and\ \bibinfo {author}
  {\bibfnamefont {J.~M.}\ \bibnamefont {Martinis}},\ }\href {\doibase
  10.1103/PhysRevLett.116.020501} {\bibfield  {journal} {\bibinfo  {journal}
  {Phys. Rev. Lett.}\ }\textbf {\bibinfo {volume} {116}},\ \bibinfo {pages}
  {020501} (\bibinfo {year} {2016})}\BibitemShut {NoStop}%
\bibitem [{\citenamefont {Chen}\ and\ \citenamefont {Xue}(2018)}]{Tao.18}%
  \BibitemOpen
  \bibfield  {author} {\bibinfo {author} {\bibfnamefont {T.}~\bibnamefont
  {Chen}}\ and\ \bibinfo {author} {\bibfnamefont {Z.-Y.}\ \bibnamefont {Xue}},\
  }\href {https://link.aps.org/doi/10.1103/PhysRevApplied.10.054051} {\bibfield
   {journal} {\bibinfo  {journal} {Phys. Rev. Appl.}\ }\textbf {\bibinfo
  {volume} {10}},\ \bibinfo {pages} {054051} (\bibinfo {year}
  {2018})}\BibitemShut {NoStop}%
\bibitem [{\citenamefont {Samkharadze}\ \emph {et~al.}(2016)\citenamefont
  {Samkharadze}, \citenamefont {Bruno}, \citenamefont {Scarlino}, \citenamefont
  {Zheng}, \citenamefont {DiVincenzo}, \citenamefont {DiCarlo},\ and\
  \citenamefont {Vandersypen}}]{Samkharadze.16}%
  \BibitemOpen
  \bibfield  {author} {\bibinfo {author} {\bibfnamefont {N.}~\bibnamefont
  {Samkharadze}}, \bibinfo {author} {\bibfnamefont {A.}~\bibnamefont {Bruno}},
  \bibinfo {author} {\bibfnamefont {P.}~\bibnamefont {Scarlino}}, \bibinfo
  {author} {\bibfnamefont {G.}~\bibnamefont {Zheng}}, \bibinfo {author}
  {\bibfnamefont {D.~P.}\ \bibnamefont {DiVincenzo}}, \bibinfo {author}
  {\bibfnamefont {L.}~\bibnamefont {DiCarlo}}, \ and\ \bibinfo {author}
  {\bibfnamefont {L.~M.~K.}\ \bibnamefont {Vandersypen}},\ }\href {\doibase
  10.1103/PhysRevApplied.5.044004} {\bibfield  {journal} {\bibinfo  {journal}
  {Phys. Rev. Appl.}\ }\textbf {\bibinfo {volume} {5}},\ \bibinfo {pages}
  {044004} (\bibinfo {year} {2016})}\BibitemShut {NoStop}%
\bibitem [{\citenamefont {Zhao}\ \emph {et~al.}(2020)\citenamefont {Zhao},
  \citenamefont {Xu}, \citenamefont {Lan}, \citenamefont {Chu}, \citenamefont
  {Tan}, \citenamefont {Yu},\ and\ \citenamefont {Yu}}]{Zhao.20}%
  \BibitemOpen
  \bibfield  {author} {\bibinfo {author} {\bibfnamefont {P.}~\bibnamefont
  {Zhao}}, \bibinfo {author} {\bibfnamefont {P.}~\bibnamefont {Xu}}, \bibinfo
  {author} {\bibfnamefont {D.}~\bibnamefont {Lan}}, \bibinfo {author}
  {\bibfnamefont {J.}~\bibnamefont {Chu}}, \bibinfo {author} {\bibfnamefont
  {X.}~\bibnamefont {Tan}}, \bibinfo {author} {\bibfnamefont {H.}~\bibnamefont
  {Yu}}, \ and\ \bibinfo {author} {\bibfnamefont {Y.}~\bibnamefont {Yu}},\
  }\href {\doibase 10.1103/PhysRevLett.125.200503} {\bibfield  {journal}
  {\bibinfo  {journal} {Phys. Rev. Lett.}\ }\textbf {\bibinfo {volume} {125}},\
  \bibinfo {pages} {200503} (\bibinfo {year} {2020})}\BibitemShut {NoStop}%
\end{thebibliography}
\end{document}